\documentclass{class_nature}
\usepackage{booktabs} 
\usepackage[table]{xcolor}
\usepackage{adjustbox}  
\usepackage{fontawesome5}
\usepackage{algorithm}
\usepackage{algorithmic}
\usepackage[T1]{fontenc}
\usepackage{textcomp}
\usepackage{gensymb}
\usepackage{amsmath} 
\usepackage{color}
\usepackage{graphicx}
\usepackage{xfrac}
\usepackage{soul} 
\usepackage{lineno} 
\usepackage[english]{babel}
\usepackage{blindtext}
\usepackage{amssymb}
\usepackage{caption}
\usepackage{array}
\usepackage{makecell}
\usepackage{comment}
\usepackage{float}
\usepackage{hyperref}
\usepackage{url}
\usepackage{siunitx}

\hypersetup{
    colorlinks=false,
    linkcolor=blue,
    filecolor=magenta,      
    urlcolor=cyan,
    pdftitle={Overleaf Example},
    pdfpagemode=FullScreen,
    }

\graphicspath{ {./Figures/} }

\makeatletter
\DeclareRobustCommand\bfseriesitshape{%
  \not@math@alphabet\itshapebfseries\relax
  \fontseries\bfdefault
  \fontshape\itdefault
  \selectfont
}
\makeatother

\renewcommand{\beginsupplement}{%
        \setcounter{table}{0}
        \renewcommand{\thetable}{S\arabic{table}}%
        \setcounter{figure}{0}
        \renewcommand{\thefigure}{S\arabic{figure}}%
     }

\title{Efficient Optimization Accelerator Framework for Multi-state Spin Ising Problems} 

\author{Chirag Garg$^{1}$ \faEnvelope[regular], Sayeef Salahuddin$^{1}$ \text{\faEnvelope[regular]}}

\begin{document}

\maketitle

\begin{affiliations}
 \item Department of Electrical Engineering and Computer Sciences, University of California, Berkeley, CA 94720, USA \\
 {\noindent \faEnvelope[regular] chirag$\_$garg@berkeley.edu; sayeef@berkeley.edu}

\end{affiliations}

 
\begin{abstract}

Ising Machines are emerging hardware architectures that efficiently solve NP-Hard combinatorial optimization problems. 
Generally, combinatorial problems are transformed into quadratic unconstrained binary optimization (QUBO) form, but this transformation often complicates the solution landscape, degrading performance, especially for multi-state problems. 
To address this challenge, we model spin interactions as generalized boolean logic function to significantly reduce the exploration space. 
We demonstrate the effectiveness of our approach on graph coloring problem using probabilistic Ising solvers, achieving similar accuracy compared to state-of-the-art heuristics and machine learning algorithms.
It also shows significant improvement over state-of-the-art QUBO-based Ising solvers, including probabilistic Ising and simulated bifurcation machines. 
We also design 1024-neuron all-to-all connected probabilistic Ising accelerator on FPGA with the proposed approach that shows $\sim$10000$\times$ performance acceleration compared to GPU-based Tabucol heuristics and reducing physical neurons by 1.5–4$\times$ over baseline Ising frameworks. 
Thus, this work establishes superior efficiency, scalability and solution quality for multi-state optimization problems.

\end{abstract}

\section*{Introduction}
New computing paradigms are getting significant attention due to exponentially growing computing needs \cite{compute_trends}. A wide variety of problems falls into the class of non-deterministic polynomial-time hard (NP-hard) and are difficult to solve optimally using conventional computing solutions \cite{simulated_annealing_kirkpatrick, np_hard_optimization}. The solution space grows exponentially with problem size, therefore making brute force searching impractical for large problem instances. However, heuristics \cite{heuristics_for_optimization} and annealing \cite{simulated_annealing_kirkpatrick} based approaches have been conventionally used to tackle these computationally hard problems in domains such as logistics \cite{logistics_ref, TSP_Dac_alberta}, biology \cite{lidar_bio_application}, integrated circuits design \cite{fpga_placement}, etc. In this context, Ising machines-based accelerators \cite{Patel2022LogicallyFactorization, supriyo2019integer, aadit2022massively, saavan2024pass, peter_coherent_ising_machine, moy2022coupled, chowdhury2023accelerated, Lo2023AnArchitecture} are currently being leveraged to efficiently find the solution to hard optimization problems. Various technologies and hardware architectures have been explored to build these Ising accelerators including quantum annealing with superconducting qubits \cite{dwaveadvantage2}, classical annealing in memristor or RRAM \cite{scalable_uim, rram_ising}, coherent Ising machines employing optical oscillators \cite{peter_coherent_ising_machine, coherent_ising_inagaki}, coupled oscillators \cite{moy2022coupled, Lo2023AnArchitecture, tianshi_oim}, neuromorphic hardware \cite{neuromorphic_ising} and stochastic circuits (probabilistic bits) \cite{saavan2024pass, Patel2022LogicallyFactorization, kerem_training, supriyo2019integer}. Quantum annealers show promising results \cite{ibm_dwave_comp} but require low temperatures making them expensive in cost and power. Therefore, classical alternatives have gained attention due to their room-temperature operation and realization using the current semiconductor process flow. This work particularly focuses on stochastic/probabilistic Ising machines to efficiently solve combinatorial optimization problems. 

\begin{figure*}

\begin{centering}
\includegraphics[width=\linewidth]{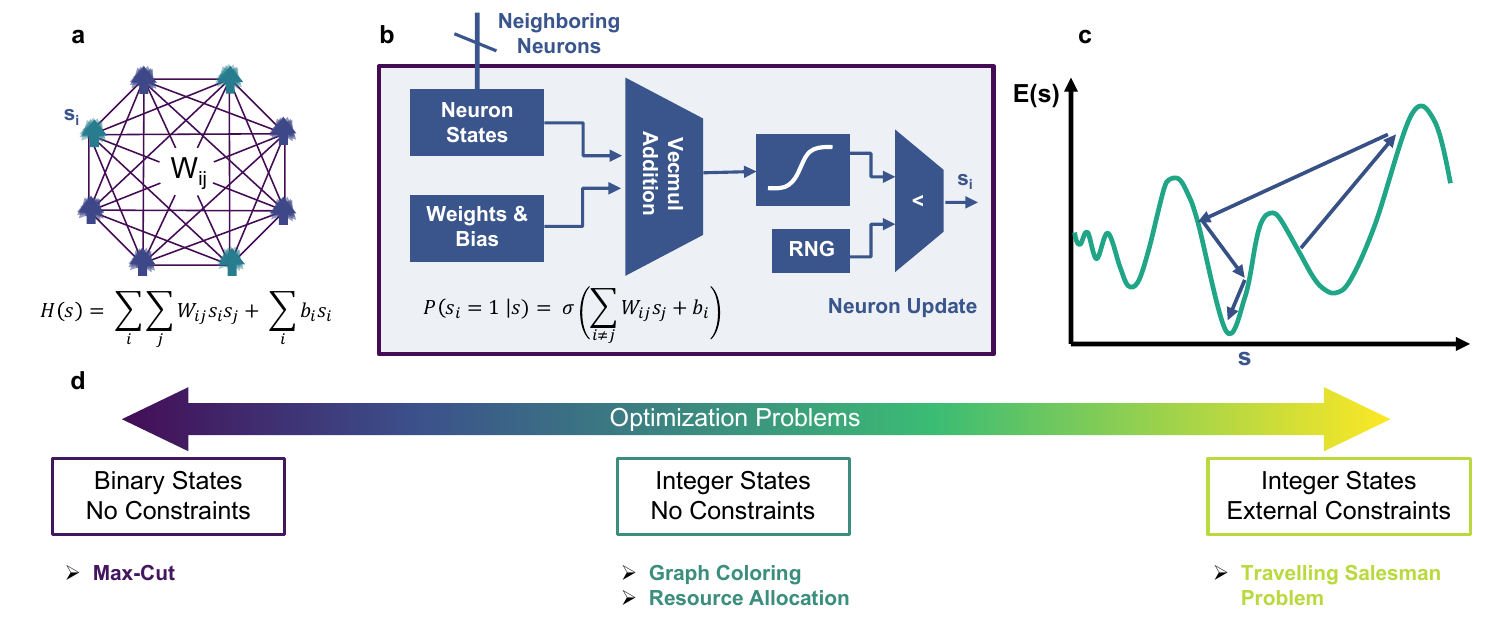}
\par\end{centering}
\caption{\label{fig:intro_fig}. \textbf{Probabilistic Ising Machine Architecture and Optimization Problem Classification} \\ 
{\textbf{a,}} Quadratic unconstrained binary optimization (QUBO) problems are represented as an Ising network with nodes representing the states and edges depicting interactions between them. 
{\textbf{b,}} Probabilistic Ising architecture based on the Boltzmann machine update rule to minimize the problem Hamiltonian and energy function.
{\textbf{c,}} The State of the Ising machine evolves and converges to the optimal energy solution.
{\textbf{d,}} Classification of optimization problems based on the state values and problem constraints. 
}
\end{figure*}

Probabilistic Ising architectures follow Boltzmann machine binary neural network principles \cite{p_bits_kerem_inv_logic, Patel2020IsingMachine}. These architectures are physically constructed using binary stochastic neurons interacting with each other and aim to minimize the Ising Hamiltonian (Fig. \ref{fig:intro_fig}a, Supplementary Note 1). The neuron update follows sigmoidal activation (Fig. \ref{fig:intro_fig}b) such that neurons stochastically move towards the minimum energy states (Fig. \ref{fig:intro_fig}c). To harness the efficient solution exploration of these hardware architectures, the cost/energy functions of many complex optimization problems are converted into Ising Hamiltonian form. Based on this conversion, these optimization problems can be broadly divided into three categories (Fig. \ref{fig:intro_fig}d) \cite{lucas_mapping}. The problem class comprised of binary solution state space with no imposed constraints has been widely mapped and efficiently solved on Ising machines \cite{Patel2020IsingMachine, coherent_ising_inagaki, Lo2023AnArchitecture}. On the other hand, problems with integer/multi-state valued solutions do not naturally convert to binary solution state space, thus leading to additional constraints in the converted Ising Hamiltonian \cite{lucas_mapping, silva2020mapping_gc_qa, constraint_graph_coloring}. This direct conversion method frequently causes the Ising machines to explore infeasible solution space,  resulting in inefficiencies \cite{scalable_uim, inaba2022potts, whitehead2023cmos_potts}. One such example is solving the graph coloring problem with Ising machines.

Graph coloring is an NP-hard optimization problem that seeks to assign different colors to the connected nodes of a graph network. It is an example of an integer optimization problem where each integral value represents the color of a node. Previously, annealing and heuristic methods (Tabu-search \cite{gc_tabucol}) were utilized to tackle this problem. These approaches achieve good solution accuracy but suffer from long runtimes for large and densely connected problem instances. Recently, learning-based approaches\cite{physics_rev_gc_gnn, li2022rethinking_gnn_gc} especially graph neural networks are applied to solve the problem accurately and efficiently at a scale. Existing Ising hardware or its modified versions are also proposed to solve this problem. However, they significantly lag behind heuristics and learning-based approaches in solution quality. To an extent, earlier works \cite{scalable_uim, inaba2022potts, whitehead2023cmos_potts} overcome this bottleneck by adopting post-processing techniques using additional hardware and software, which adversely affects area and computation time.

In this work, we present an end-to-end probabilistic Ising implementation that combines advances in multi-state problem mapping, spin interaction design, and an efficient hardware architecture to demonstrate significant improvement in area, solution accuracy, and time-to-solution. First, we propose vectorized mapping that represents the node colors as binary vectors rather than using the customary one-hot form. This circumvents the additional mapping constraint in the Ising Hamiltonian that arises due to one-hot encoding \cite{lucas_mapping}. Thus, it completely discards the exploration of infeasible/invalid solution space and improves the solution quality. Second, the interactions among the binary vector states are modeled using truth tables employed in digital logic. Third, we implement an efficient FPGA accelerator that uses higher-order multiplexers to represent these truth tables. Altogether, this multi-state Ising implementation shows approximately 100000$\times$ speed improvement and 5$\times$ power improvement compared to its GPU-based implementation for solving problems up to 256 nodes and 16 colors graph coloring problems. This current methodology is then integrated with the parallel tempering approach to improve the solution quality and provide competitive or even better coloring results than Tabucol heuristics and machine learning algorithms.

\begin{figure*}
\begin{centering}
\includegraphics[width=\linewidth]{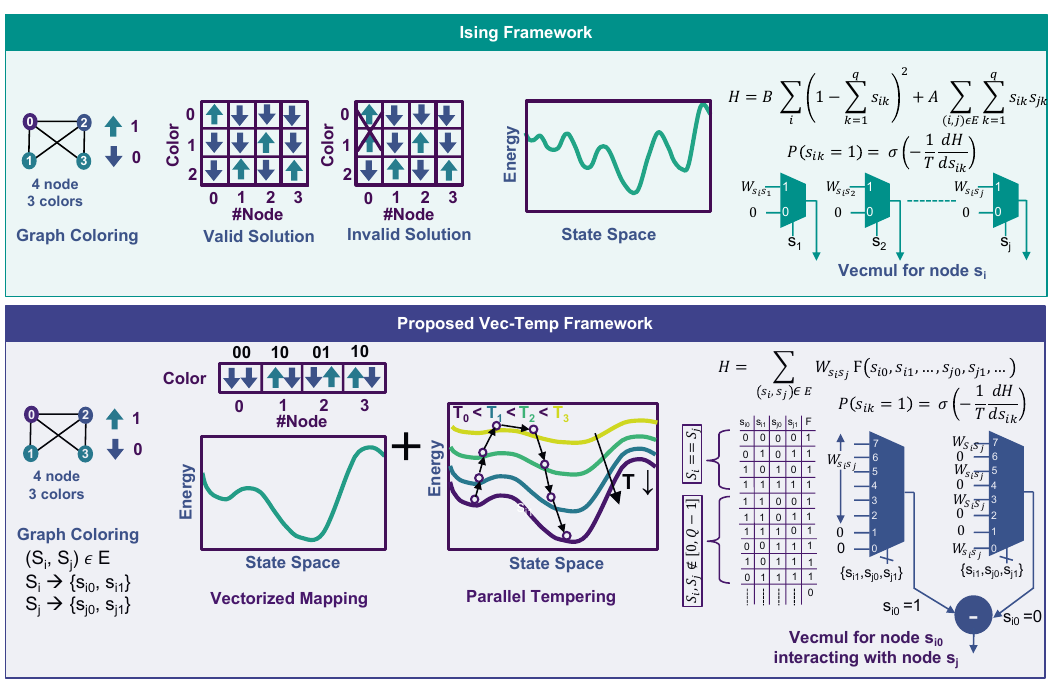}
\par\end{centering}
\caption{\label{fig:mapping_comp} \textbf{Comparison between Ising and Vectorized Mapping Approaches} \\ 
{\textbf{a,}} Existing Ising framework employs one-hot encoding to map $N$ node $q$ color graph coloring problem to $Nq$ nodes. The encoding constraint is enforced by adding an extra term in the Hamiltonian, creating a complex optimization landscape that can result in suboptimal ground states. This mapping preserves QUBO Hamiltonian, therefore, the interactions are modeled via binary multiplexers in hardware. {\textbf{b,}} Vectorized framework maps the $q$ color state of a node using 
$\lceil log2q \rceil$ hardware nodes. It demonstrates the proposed mapping framework applied for 4 nodes and 3 colors graph coloring problem. This framework searches only valid solutions, helping the hardware reach ground states more easily.
Parallel tempering further improves solution exploration. This mapping can be modeled in a truth table format, which is directly mapped to a higher-order multiplexer in hardware.}

\end{figure*}

\begin{results}
\section{Ising Mapping Framework}

A wide range of combinatorial optimization problems belongs to multistate/integer-valued solution space, including graph coloring, knapsack problems \cite{koopmans1957assignment_knapsack}, etc. Graph coloring problem, investigated in this work, has applications, such as layout decomposition\cite{layout_decomposition}, register allocation\cite{register_allocation}, logic minimization\cite{logic_minimization}, scheduling\cite{physics_rev_gc_gnn}, and many more. The problem instance that aims to color $N$ nodes with $q$ colors, requires $Nq$ physical neurons/nodes in the Ising framework. Because this framework relies on one-hot encoding to represent node colors, additional constraints are imposed in problem Hamiltonian ($H_A$ in Eq. \ref{eq:graph_coloring}) to enforce the legal combination of one-hot bits \cite{lucas_mapping} (represented by Eq. \ref{eq:coloring_constraint}). Therefore, the Ising Hamiltonian (H) for graph coloring problem becomes $H_A + H_B$.
\begin{equation}
    \label{eq:graph_coloring}
    H_{A} = A\sum_{i, j \in E} \sum_{k=1}^{q}s_{ik}s_{jk}  \\ 
\end{equation}
\begin{equation}
    \label{eq:coloring_constraint}
    H_{B} = B\sum_{i}(1-\sum_{k=1}^{q}s_{ik})^{2} \\ 
\end{equation}
In the equations, $s_{ik}$ represents $k^{th}$ one-hot bit for node $i$, $E$ denotes the set of all edges in the problem graph, $A$ is the connectivity weighting factor, and $B$ is the one-hot constraint factor. This soft constraint Hamiltonian is expected to be zero for the ground state energy solution. 
However, it often takes the non-zero value, thereby making Ising samplers inefficient in solving such problems. (see Fig. \ref{fig:mapping_comp}a). The one-hot encoding requires $q$ bits to represent $q$ possible colors of a node. As a result, Ising machines search over $2^{q}$ states with only $q$ valid states in the solution space. It makes the optimization hard and deteriorates the solution quality. Hence, the Ising frameworks following one-hot encoding often fail to solve this class of problems. Despite these challenges, this approach is often adopted for most multistate optimization problems because it leads to quadratic interactions supported in most Ising machines \cite{Patel2020IsingMachine,supriyo2019integer, Lo2023AnArchitecture, coherent_ising_inagaki}. In probabilistic computing \cite{saavan2024pass}, binary multiplexers offer simplistic and efficient realization for state-weight multiplication. 

\section{Vectorized Mapping Framework}

In this work, we propose a vectorized mapping approach (Fig. \ref{fig:mapping_comp}b) to tailor the graph coloring problem using a binary encoding technique. It represents the $q$ color state using the $\lceil log2q \rceil$ ($=n$) bit vector, which means color for node $S_i$ is represented as \{$s_{i0}, s_{i1}, \ldots, s_{i(n-1)}$\}. Therefore, this mapping requires $N\lceil log2q \rceil$ physical neurons/nodes for a graph coloring problem with $N$ nodes and $q$ colors. In this way, it eliminates all invalid state space and removes the extra constraints in Ising Hamiltonian (Eq. \ref{eq:coloring_constraint}). Therefore, the proposed approach leads to a solvable energy landscape without additional complexities. We also adopt parallel tempering  \cite{earl2005paralleltempering} to enhance the solution exploration and improve the solution quality. 

\begin{equation}
    \label{eq:vectorized_hamiltonian}
    H = \sum_{(S_i S_j)\in E} W_{S_i S_j}F(s_{i0}, s_{i1}, \ldots, s_{i(n-1)},  s_{j0}, s_{j1}, \ldots, s_{j(n-1)}) \\ 
\end{equation}

The proposed vector mapping is a generic approach that can be applied to any multistate optimization problem being solved on Ising accelerators (see Fig. \ref{fig:mapping_comp}b). This approach represents the states in binary representation and only requires modeling the function operator $F$ in the Hamiltonian H ( Eq. \ref{eq:vectorized_hamiltonian}) for a specific problem. In  Eq. \ref{eq:vectorized_hamiltonian} , $W_{S_iS_j}$ represents the edge (E) weight connecting the nodes $S_i$ and $S_j$, and function operator $F$ is modeled in a truth table-based format. Algorithm \ref{Algorithm_1} describes the formulation of the truth table for the graph coloring problem and Fig. \ref{fig:mapping_comp} shows the formulation of the truth table of a 4 nodes, 3 colors graph coloring problem. In the cases when nodes ($S_{i}$ and $S_{j}$) take the same value or color value greater than q (invalid colors), $F$ becomes one. $F$ is zero for the rest of the cases. The resulting Hamiltonian is a higher-order polynomial due to the $F$ operator function. Ideally, graph coloring criteria are met when the Hamiltonian energy reaches its global minimum of zero, which corresponds to the condition where 
$F$ is equal to zero.
Further, the truth table-based algorithm facilitates its representation as multiplexers, particularly in the case of Ising hardware, which digitally models spin interactions. In this work, we leverage the probabilistic Ising machines framework to develop the accelerator following the proposed vectorized mapping. The hardware architecture for this scheme is derived from the standard node update rule for probabilistic Ising machines. For $k^{th}$ bit of node $S_{i}$, the update rule is as follows: 
\begin{equation}
    \label{eq:update_rule}
    P(s_{ik}=1) = \sigma\left(-\frac{1}{T}\frac{dH}{ds_{ik}}\right) \\ 
\end{equation}
where $s_{ik}$ takes binary value 0 and 1, $T$  is temperature coefficient and $\sigma$ is sigmoid function ($=1/(1+e^x))$ for input value $x$). The binary nature of $s_{ik}$ converts the differentiation transforms into $\Delta H$ equal to $H_{s_{ik}=1}$ - $H_{s_{ik}=0}$. This $\Delta H$ is implemented in hardware using higher-order multiplexers followed by a subtraction unit (see Fig. \ref{fig:mapping_comp}), called as vecmul unit. The rest of the hardware implementation remains the same as shown in Fig. \ref{fig:intro_fig}. 

\section{Accuracy Analysis and Benchmarks}

Next we investigate the proposed vectorized mapping approach for the graph coloring problem. For benchmarking, we have used it on publicly available COLOR dataset \cite{trick2002color_dataset}. First, we implement probabilistic Ising frameworks following one-hot encoding (Ising) and proposed binary encoding (vectorized) approaches on GPU for comparison. Fig. \ref{fig:solution_exploration} shows the results for the $queen13\_13$ problem, which is deemed a {\emph hard} problem \cite{physics_rev_gc_gnn}. For both schemes, the exploration for solution happens in a similar way (Fig. \ref{fig:solution_exploration}a). 
Fig. \ref{fig:solution_exploration}b illustrates that the vectorized mapping gives superior coloring results as compared to Ising framework. Even with an optimal set of connectivity weight factor and one-hot constraint weight factor parameters (see Supplementary Fig. \ref{fig:hyperparam}), the effectiveness of the Ising framework is limited by the one-hot encoding constraint in Eq.\ref{eq:coloring_constraint}.
This comparison for other problem instances in the dataset is shown in the Supplementary Fig. \ref{fig:incorr_edge} and \ref{fig:energy_evolution}. Further, we employ the parallel tempering method to enhance the exploration while sampling the energy landscape. This method involves the simultaneous simulation of M replica systems with vectorized mapping running at different temperature schedules. At high temperatures, the system tends to explore the broad region of the solution space while low temperatures allow precise sampling around a particular region (Fig. \ref{fig:mapping_comp}b). High-temperature replicas ensure that the system doesn't get stuck around the local minimum solution. The states of higher-temperature replica systems are exchanged with low-temperature systems via a swap rule (see Methods Section 1). This state exchange phenomenon enables the exploration of certain parts of solution space which could not be possible via a single temperature schedule. Fig. \ref{fig:solution_exploration}c depicts the energy exploration for the $queen13\_13$ problem using the parallel tempering approach. It clearly shows that the system continuously escapes from the minimum local energy solution and therefore, achieves the best possible solution accuracy. 


We evaluate the accuracy of vectorized mapping solutions against learning-based methods (graph neural network \cite{li2022rethinking_gnn_gc}   and its GraphSage architecture PI-SAGE \cite{physics_rev_gc_gnn} ), heuristic approaches (Tabucol \cite{gc_tabucol, Lewis2021_gc_book}), and state-of-the-art Ising solvers (probabilistic Ising machines \cite{all_to_kerem_benchmark}, simulated bifurcation\cite{sbm_science, sbm_git} ) for graph coloring across standard benchmark datasets. GNN-based solvers frame the coloring problem as a multi-class node classification problem and employ unsupervised training by formulating a loss function. In contrast, the Tabucol technique searches the ground state energy solution by moving small steps and maintains a tabu list to avoid cycling around local minima.
 We run Tabucol heuristics, probabilistic Ising machines, and vectorized mapping framework with single-flip Gibbs sampling on Nvidia A100 Tensor Core GPU and solve the same problems 200 times with 1000 iteration (update) steps each. Additionally, simulated bifurcation machines algorithm has been run on the same GPU solving  the same problems 200 times with 10000 iteration steps each. We report best possible results achieved by the described methods and compares them  in Table \ref{fig:accuracy_table} and \ref{fig:accuracy_table_citation}. It includes easy, medium, and hard problem instances labeled in work \cite{dsatur}. The Ising approach performs worse by only being able to solve small and easy problem instances accurately. Among GNN-based solvers, the GraphSage architecture (PI-SAGE) offers better solution accuracy but it suffers from longer training times \cite{physics_rev_gc_gnn}. By contrast, proposed vectorized mapping gives competitive coloring results compared to Tabucol heuristics and PI-SAGE GNN while having slightly lower accuracy for hard-to-solve problem instances. Employing the described parallel tempering with vectorized mapping reduces the error up to 50\% on these {\emph hard} problems and therefore performs better than other methods. 

\section{Evaluation of Success Probability and Time-to-Solution}

We employ the success probability metric extensively used in other works\cite{Lo2023AnArchitecture, Hamerly2019ExperimentalAnnealer} that captures the solution quality of statistical algorithms. We define error as number of incorrectly colored edges divided by total edges in the problem graph. To calculate success probability ($p_s$), we run each coloring problem for 200 times (or parallel runs) using each algorithm and calculate the success in getting the error less than 2$\%$ for those iterations. The results in Fig. \ref{fig:area_TTS}b confirms that the vectorized mapping achieves competitive success probability against the Tabucol heuristic approach and produces a better quality solution compared to state-of-the-art Ising solvers, including probabilistic Ising machine and Simulated Bifurcation machines. Additionally,  time-to-solution (TTS) in Eq. \ref{eq:TTS} is defined as the time needed to obtain a solution within a specified accuracy across multiple runs, with a probability of 99$\%$. $T_{comp}$ represents the average time to complete a single run. The algorithms that achieve success probability greater than 99$\%$, TTS is defined as the average time to reach the solution across parallel runs. Using this methodology, Fig. \ref{fig:area_TTS}c reports the TTS for different statistical algorithms used to solve the graph coloring problem instances. Overall, the success probability and TTS metric capture a critical aspect of solver performance. A higher success probability reflects a solver's ability to tackle problem instances more efficiently, requiring fewer attempts. This efficiency is represented by incorporating a factor in TTS formulation that accounts for the influence of success probability, providing a robust evaluation of solver effectiveness in terms of both accuracy and computational efficiency.

\begin{equation}
    \label{eq:TTS}
    TTS = T_{comp}*\frac{log(1-0.99)}{log(1-p_{s})} \\ 
\end{equation}

\section{FPGA Accelerator Implementation and Results}

We implement the proposed vectorized architecture (Fig. \ref{fig:mapping_comp}b) onto VCU118 FPGA to establish the performance acceleration and energy efficiency benchmarks. The FPGA accelerator uses the memory-mapped IO interface used by software applications to program the problem weights and take the solution out (see Methods Section 2 for more details). The accelerator (Fig. \ref{fig:intro_fig}) fetches the weight data from the memory and does neuron states-weight multiplication. This process includes computing the F-operator, defined in Algorithm \ref{Algorithm_1} which is implemented in hardware using truth tables or higher-order multiplexers, as illustrated in Fig. \ref{fig:mapping_comp}b. The structure of this operator is largely determined by the graph's coloring constraints. The accumulated product, represented with 8-bit precision, is passed through a sigmoid activation function that is implemented using a look-up table (LUT) containing $2^8$ (256) entries. The LUT generates a 16-bit output, which is then compared with a 16-bit random number generated using a Linear Feedback Shift Register (LFSR)  to get the updated node value. The architecture runs at 90 MHz clock frequency where only one node gets updated in each cycle. Further, it supports graph coloring problem sizes up to 256 nodes all-to-all connected with a maximum chromatic number of 16, equivalent to 1024 vectorized nodes in total. The framework achieves the same accuracy for dataset problems as the GPU-based vectorized mapping (see Table \ref{fig:accuracy_table}).  

The proposed architecture not only gives better accuracy than the baseline Ising-based architecture following one-hot encoding but also reduces the area implementation. The area implementation is quantified in terms of number of physical nodes required to map any problem instance and solve it. The increase in physical nodes leads to an increase in the number of interactions and hence accumulator size grows leading to a penalty on the implementation area. Fig. \ref{fig:area_TTS}a shows that the vectorized mapping requires 1.5-4 times fewer nodes for the graph coloring problem instances with chromatic numbers up to 16. 

The FPGA architecture also uses single flip-Gibbs sampling to update the node, therefore, N$\left\lceil \log_{2}(Q) \right\rceil$ nodes need to be updated for vectorized mapping system for one complete iteration or time step. Thus, the time complexity for the proposed framework will scale with $O$($N\left\lceil \log_{2}(Q) \right\rceil$). The time update for a single node update affects the prefactor of thes mentioned time complexity. Specifically, in probabilistic Ising/vectorized, this prefactor is contribution of multiplexer (states-weight multiplication), accumulator, sigmoid activation calculation and comparator delays. Owing to binary nature of nodes, the accumulator dominates the node update timing\cite{Patel2022LogicallyFactorization}. As a result, introducing higher-order multiplexers has minimal impact on the overall time complexity scaling.

The FPGA implementation of the proposed vectorized mapping approach shows $\sim$10000$\times$  speedup compared to Tabucol heuristics and Ising solvers (Probabilistic Ising and Simulated Bifurcation) implementation on Nvidia A100 Tensor Core GPU (see Fig.  \ref{fig:area_TTS}c). Vectorized mapping on GPU could not take advantage of efficient multiplication of binary states with weights and, provides only comparable time-to-solution to heuristics.
This same trend is expected to hold for any accelerators that are is built to support these binary calculations. 
Similarly, Ising mapping-based solvers, including  Probabilistic Ising and Simulated Bifurcation, can benefit from FPGA acceleration \cite{all_to_kerem_benchmark, fpga_sbm}, potentially narrowing the performance gap with our FPGA implementation of the vectorized mapping. Nevertheless, for problems with more than 100 nodes, these solvers may continue to yield sub-optimal solutions.
The FPGA accelerator also offers $\sim$5$\times$ power improvements over GPU-based vectorized mapping implementation. Therefore, accelerating the time-to-solution by > 4 orders of magnitude on FPGA at a low power budget significantly improves energy efficiency Fig.  \ref{fig:area_TTS}d. 

One question that may arise is: is the acceleration achieved for the proposed method over Heuristics solely due to the fact that the vectorized method was implemented on FPGA? In this regard, we note here that the Heuristics take advantage of sequential strategies that do not scale well on parallel architecture. This has been studied extensively in the literature. For example, in  Supplementary Fig. \ref{fig:cpu_gpu_tabu_vec}a we show a comparison between CPU and GPU implementations of Tabucol. It is clearly seen that the GPU implementation provides virtually no acceleration. By contrast GPU implementation of the proposed vectorized method shows large acceleration as shown in  Supplementary Fig. \ref{fig:cpu_gpu_tabu_vec}b. This underscores a strength of the proposed method that makes it amenable for scaling on specialized hardware such as the FPGA.
\end{results}
\section*{Discussion}

Current methods employ the one-hot state encoding framework to map and solve multistate spin Ising problems onto Ising machines. This work presents an alternative approach of vectorized mapping that maps the state using binary encoding. It not only reduces state exploration from $2^{qN}$ to $2^{\lceil log2q \rceil N}$ for a q color problem instance but also removes the constraints resulting from the constraint-heavy Ising mapping. As a result, the proposed approach converges close to the optimal solution achieved by heuristics and learning based approaches. Existing works \cite{Lo2023AnArchitecture, saavan2024pass, coherent_ising_inagaki} on Ising Machines and its derivative frameworks generally aims to achieve accuracy within a certain error range compared to heuristics. However, we show that our method combined with parallel tempering achieves the solution accuracy even better than the heuristics for some of the problems, while retaining the ability to significant speed up on a custom accelerator. We also present a generalized truth table-based method that can be leveraged to map other multi-state problems. Using probabilistic Ising machine architectures, these truth tables to model the neuron interactions can be directly mapped onto higher order multiplexers. We have implemented this architecture on FPGA to benchmark time-to-solution, energy efficiency, and area efficiency. Overall, the hardware implementation consumes 5W power and achieves approximately $\sim$10000$\times$ speedup compared to Tabucol heuristics and $\sim$100000$\times$ compared to vectorized mapping on GPU.  The presented hardware and software framework provides a new way to substantially expand the capabilities of the Ising machines to accurately handle a wide range of multistate optimization problems in a performance and energy-efficient manner.

\clearpage
\newpage
\begin{methods}
\section*{1. Parallel Tempering}

The parallel tempering algorithm utilizes multiple Markov chains (replica chains) running at different temperature schedules representing different probability distributions. It allows a broader exploration of the energy landscape and facilitates better mixing to avoid local energy minima solutions. In this work, we implement 100 parallel chains running at a constant temperature geometrically spaced from temp0 (0.01) and tempn (40). After every 15 sampling steps, adjacent pairs of chains are swapped alternatively with odd-leading and even-leading chain indexes. The update rule is given as:

\begin{equation}
    \label{eq:pt_swap_rule}
    r = exp\left[\left(\frac{1}{T_{1}}-\frac{1}{T_{2}}\right)(H(s_{T1})-H(s_{T2}))\right] \\ 
\end{equation}

\begin{equation}
    \label{eq:pt_swap_rule2}
    P_{swap}(s_{T1}-s_{T2}) = min(1, r) \\ 
\end{equation}

\section*{2. FPGA Setup PCIe}
In this work, we implement the vectorized mapping on the Xilinx Virtex UltraScale+ VCU118 FPGA evaluation kit. The FPGA is interfaced with the CPU using 
peripheral component interconnect express (PCIe) interface. In particular, we use an open-source xillybus IP core for the interface with data transfer capabilities of 50MB/s. The data is transferred via a memory-mapped interface implemented using block memory and a designed memory controller. 

The digital implementation of 256 nodes and 16 color accelerators supports a network of 1024 probabilistic Ising nodes. These nodes consist of an LFSR-based pseudorandom number generator with a fixed seed, a lookup table-based sigmoidal activation, higher order multiplexer-based matrix multiplication, and a threshold that generates the output state of the neuron. These states are stored in the local memory of FPGA and then transferred to CPU via the PCIe interface. 

\end{methods}

\clearpage
\newpage
\begin{addendum}

\item [Data Availability]
All processed data generated in this study are provided in the main text. The data supporting the plots in this paper can be found in the \href{https://github.com/chiraggarg24/Optimization_Integer_Ising.git} {GitHub} repository \cite{git_data}. Other findings of this study are available from the corresponding authors upon request.

\item [Code Availability]
Code will be made available on reasonable request by emailing C.G. or S.S.

\end{addendum}

\bibliographystyle{naturemag}
\bibliography{references.bib}

\begin{thebibliography}{10}
\expandafter\ifx\csname url\endcsname\relax
  \def\url#1{\texttt{#1}}\fi
\expandafter\ifx\csname urlprefix\endcsname\relax\def\urlprefix{URL }\fi
\providecommand{\bibinfo}[2]{#2}
\providecommand{\eprint}[2][]{\url{#2}}

\bibitem{compute_trends}
\bibinfo{author}{Sevilla, J.} \emph{et~al.}
\newblock \bibinfo{title}{Compute trends across three eras of machine learning}.
\newblock In \emph{\bibinfo{booktitle}{2022 International Joint Conference on Neural Networks (IJCNN)}}, \bibinfo{pages}{1--8} (\bibinfo{year}{2022}).

\bibitem{simulated_annealing_kirkpatrick}
\bibinfo{author}{Kirkpatrick, S.}, \bibinfo{author}{Gelatt, C.~D.} \& \bibinfo{author}{Vecchi, M.~P.}
\newblock \bibinfo{title}{Optimization by simulated annealing}.
\newblock \emph{\bibinfo{journal}{Science}} \textbf{\bibinfo{volume}{220}}, \bibinfo{pages}{671--680} (\bibinfo{year}{1983}).
\newblock \eprint{https://www.science.org/doi/pdf/10.1126/science.220.4598.671}.

\bibitem{np_hard_optimization}
\bibinfo{author}{Hochba, D.~S.}
\newblock \bibinfo{title}{Approximation algorithms for np-hard problems}.
\newblock \emph{\bibinfo{journal}{SIGACT News}} \textbf{\bibinfo{volume}{28}}, \bibinfo{pages}{40–52} (\bibinfo{year}{1997}).

\bibitem{heuristics_for_optimization}
\bibinfo{author}{Colorni, A.} \emph{et~al.}
\newblock \bibinfo{title}{Heuristics from nature for hard combinatorial optimization problems}.
\newblock \emph{\bibinfo{journal}{International Transactions in Operational Research}} \textbf{\bibinfo{volume}{3}}, \bibinfo{pages}{1--21} (\bibinfo{year}{1996}).
\newblock \eprint{https://onlinelibrary.wiley.com/doi/pdf/10.1111/j.1475-3995.1996.tb00032.x}.

\bibitem{logistics_ref}
\bibinfo{author}{Weinberg, S.~J.}, \bibinfo{author}{Sanches, F.}, \bibinfo{author}{Ide, T.}, \bibinfo{author}{Kamiya, K.} \& \bibinfo{author}{Correll, R.}
\newblock \bibinfo{title}{Supply chain logistics with quantum and classical annealing algorithms}.
\newblock \emph{\bibinfo{journal}{Scientific Reports}} \textbf{\bibinfo{volume}{13}}, \bibinfo{pages}{4770} (\bibinfo{year}{2023}).

\bibitem{TSP_Dac_alberta}
\bibinfo{author}{Tao, Q.} \& \bibinfo{author}{Han, J.}
\newblock \bibinfo{title}{Solving traveling salesman problems via a parallel fully connected ising machine}.
\newblock In \emph{\bibinfo{booktitle}{Proceedings of the 59th ACM/IEEE Design Automation Conference}}, DAC '22, \bibinfo{pages}{1123–1128} (\bibinfo{publisher}{Association for Computing Machinery}, \bibinfo{address}{New York, NY, USA}, \bibinfo{year}{2022}).

\bibitem{lidar_bio_application}
\bibinfo{author}{Li, R.~Y.}, \bibinfo{author}{Di~Felice, R.}, \bibinfo{author}{Rohs, R.} \& \bibinfo{author}{Lidar, D.~A.}
\newblock \bibinfo{title}{Quantum annealing versus classical machine learning applied to a simplified computational biology problem}.
\newblock \emph{\bibinfo{journal}{npj Quantum Information}} \textbf{\bibinfo{volume}{4}}, \bibinfo{pages}{14} (\bibinfo{year}{2018}).

\bibitem{fpga_placement}
\bibinfo{author}{Gerlach, T.} \emph{et~al.}
\newblock \bibinfo{title}{Fpga-placement via quantum annealing}.
\newblock In \emph{\bibinfo{booktitle}{Proceedings of the 2024 ACM/SIGDA International Symposium on Field Programmable Gate Arrays}}, FPGA '24, \bibinfo{pages}{43} (\bibinfo{publisher}{Association for Computing Machinery}, \bibinfo{address}{New York, NY, USA}, \bibinfo{year}{2024}).

\bibitem{Patel2022LogicallyFactorization}
\bibinfo{author}{Patel, S.}, \bibinfo{author}{Canoza, P.} \& \bibinfo{author}{Salahuddin, S.}
\newblock \bibinfo{title}{{Logically synthesized and hardware-accelerated restricted Boltzmann machines for combinatorial optimization and integer factorization}}.
\newblock \emph{\bibinfo{journal}{Nature Electronics 2022 5:2}} \textbf{\bibinfo{volume}{5}}, \bibinfo{pages}{92--101} (\bibinfo{year}{2022}).

\bibitem{supriyo2019integer}
\bibinfo{author}{Borders, W.~A.} \emph{et~al.}
\newblock \bibinfo{title}{Integer factorization using stochastic magnetic tunnel junctions}.
\newblock \emph{\bibinfo{journal}{Nature}} \textbf{\bibinfo{volume}{573}}, \bibinfo{pages}{390--393} (\bibinfo{year}{2019}).

\bibitem{aadit2022massively}
\bibinfo{author}{Aadit, N.~A.} \emph{et~al.}
\newblock \bibinfo{title}{Massively parallel probabilistic computing with sparse ising machines}.
\newblock \emph{\bibinfo{journal}{Nature Electronics}} \textbf{\bibinfo{volume}{5}}, \bibinfo{pages}{460--468} (\bibinfo{year}{2022}).

\bibitem{saavan2024pass}
\bibinfo{author}{Patel, S.} \emph{et~al.}
\newblock \bibinfo{title}{{PASS: An Asynchronous Probabilistic Processor for Next Generation Intelligence}}.
\newblock \emph{\bibinfo{journal}{ArXiv}}  (\bibinfo{year}{2024}).
\newblock \eprint{/abs/2409.10325}.

\bibitem{peter_coherent_ising_machine}
\bibinfo{author}{McMahon, P.~L.} \emph{et~al.}
\newblock \bibinfo{title}{A fully programmable 100-spin coherent ising machine with all-to-all connections}.
\newblock \emph{\bibinfo{journal}{Science}} \textbf{\bibinfo{volume}{354}}, \bibinfo{pages}{614--617} (\bibinfo{year}{2016}).
\newblock \eprint{https://www.science.org/doi/pdf/10.1126/science.aah5178}.

\bibitem{moy2022coupled}
\bibinfo{author}{Moy, W.} \emph{et~al.}
\newblock \bibinfo{title}{A 1,968-node coupled ring oscillator circuit for combinatorial optimization problem solving}.
\newblock \emph{\bibinfo{journal}{Nature Electronics}} \textbf{\bibinfo{volume}{5}}, \bibinfo{pages}{310--317} (\bibinfo{year}{2022}).

\bibitem{chowdhury2023accelerated}
\bibinfo{author}{Chowdhury, S.}, \bibinfo{author}{Camsari, K.~Y.} \& \bibinfo{author}{Datta, S.}
\newblock \bibinfo{title}{Accelerated quantum monte carlo with probabilistic computers}.
\newblock \emph{\bibinfo{journal}{Communications Physics}} \textbf{\bibinfo{volume}{6}}, \bibinfo{pages}{1--7} (\bibinfo{year}{2023}).

\bibitem{Lo2023AnArchitecture}
\bibinfo{author}{Lo, H.}, \bibinfo{author}{Moy, W.}, \bibinfo{author}{Yu, H.}, \bibinfo{author}{Sapatnekar, S.} \& \bibinfo{author}{Kim, C.~H.}
\newblock \bibinfo{title}{{An Ising solver chip based on coupled ring oscillators with a 48-node all-to-all connected array architecture}}.
\newblock \emph{\bibinfo{journal}{Nature Electronics}} \textbf{\bibinfo{volume}{6}}, \bibinfo{pages}{771--778} (\bibinfo{year}{2023}).

\bibitem{dwaveadvantage2}
\bibinfo{author}{McGeoch, C.}, \bibinfo{author}{Farre, P.} \& \bibinfo{author}{Boothby, K.}
\newblock \bibinfo{title}{{The D-Wave Advantage2 Prototype}}.
\newblock \bibinfo{type}{Tech. Rep.}, \bibinfo{institution}{D-Wave Systems Inc.} (\bibinfo{year}{2023}).
\newblock \bibinfo{note}{Technical Report}.

\bibitem{scalable_uim}
\bibinfo{author}{Yue, W.}, \bibinfo{author}{Zhang, T.}, \bibinfo{author}{Jing, Z.} \emph{et~al.}
\newblock \bibinfo{title}{A scalable universal ising machine based on interaction-centric storage and compute-in-memory}.
\newblock \emph{\bibinfo{journal}{Nature Electronics}}  (\bibinfo{year}{2024}).

\bibitem{rram_ising}
\bibinfo{author}{Chiang, H.-W.}, \bibinfo{author}{Nien, C.-F.}, \bibinfo{author}{Cheng, H.-Y.} \& \bibinfo{author}{Huang, K.-P.}
\newblock \bibinfo{title}{Reaim: A reram-based adaptive ising machine for solving combinatorial optimization problems}.
\newblock In \emph{\bibinfo{booktitle}{2024 ACM/IEEE 51st Annual International Symposium on Computer Architecture (ISCA)}}, \bibinfo{pages}{58--72} (\bibinfo{year}{2024}).

\bibitem{coherent_ising_inagaki}
\bibinfo{author}{Inagaki, T.} \emph{et~al.}
\newblock \bibinfo{title}{A coherent ising machine for 2000-node optimization problems}.
\newblock \emph{\bibinfo{journal}{Science}} \textbf{\bibinfo{volume}{354}}, \bibinfo{pages}{603--606} (\bibinfo{year}{2016}).
\newblock \eprint{https://www.science.org/doi/pdf/10.1126/science.aah4243}.

\bibitem{tianshi_oim}
\bibinfo{author}{Wang, T.} \& \bibinfo{author}{Roychowdhury, J.}
\newblock \bibinfo{title}{Oim: Oscillator-based ising machines for solving combinatorial optimisation problems}.
\newblock In \bibinfo{editor}{McQuillan, I.} \& \bibinfo{editor}{Seki, S.} (eds.) \emph{\bibinfo{booktitle}{Unconventional Computation and Natural Computation}}, \bibinfo{pages}{232--256} (\bibinfo{publisher}{Springer International Publishing}, \bibinfo{address}{Cham}, \bibinfo{year}{2019}).

\bibitem{neuromorphic_ising}
\bibinfo{author}{Chen, Z.} \emph{et~al.}
\newblock \bibinfo{title}{{ON-OFF Neuromorphic ISING Machines using Fowler-Nordheim Annealers}}.
\newblock \emph{\bibinfo{journal}{ArXiv}}  (\bibinfo{year}{2024}).
\newblock \eprint{2406.05224}.

\bibitem{kerem_training}
\bibinfo{author}{Niazi, S.}, \bibinfo{author}{Chowdhury, S.}, \bibinfo{author}{Aadit, N.} \emph{et~al.}
\newblock \bibinfo{title}{Training deep boltzmann networks with sparse ising machines}.
\newblock \emph{\bibinfo{journal}{Nature Electronics}} \textbf{\bibinfo{volume}{7}}, \bibinfo{pages}{610--619} (\bibinfo{year}{2024}).

\bibitem{ibm_dwave_comp}
\bibinfo{author}{McGeoch, C.~C.}, \bibinfo{author}{Chern, K.}, \bibinfo{author}{Farr{\'e}, P.} \& \bibinfo{author}{King, A.~K.}
\newblock \bibinfo{title}{{A comment on comparing optimization on D-Wave and IBM quantum processors}}.
\newblock \emph{\bibinfo{journal}{ArXiv}}  (\bibinfo{year}{2024}).
\newblock \eprint{2406.19351}.

\bibitem{p_bits_kerem_inv_logic}
\bibinfo{author}{Camsari, K.~Y.}, \bibinfo{author}{Faria, R.}, \bibinfo{author}{Sutton, B.~M.} \& \bibinfo{author}{Datta, S.}
\newblock \bibinfo{title}{Stochastic $p$-bits for invertible logic}.
\newblock \emph{\bibinfo{journal}{Phys. Rev. X}} \textbf{\bibinfo{volume}{7}}, \bibinfo{pages}{031014} (\bibinfo{year}{2017}).

\bibitem{Patel2020IsingMachine}
\bibinfo{author}{Patel, S.}, \bibinfo{author}{Chen, L.}, \bibinfo{author}{Canoza, P.} \& \bibinfo{author}{Salahuddin, S.}
\newblock \bibinfo{title}{{Ising Model Optimization Problems on a FPGA Accelerated Restricted Boltzmann Machine}}  (\bibinfo{year}{2020}).

\bibitem{lucas_mapping}
\bibinfo{author}{Lucas, A.}
\newblock \bibinfo{title}{Ising formulations of many np problems}.
\newblock \emph{\bibinfo{journal}{Frontiers in Physics}} \textbf{\bibinfo{volume}{2}} (\bibinfo{year}{2014}).

\bibitem{silva2020mapping_gc_qa}
\bibinfo{author}{Silva, C.}, \bibinfo{author}{Aguiar, A.}, \bibinfo{author}{Lima, P.} \emph{et~al.}
\newblock \bibinfo{title}{Mapping graph coloring to quantum annealing}.
\newblock \emph{\bibinfo{journal}{Quantum Machine Intelligence}} \textbf{\bibinfo{volume}{2}}, \bibinfo{pages}{16} (\bibinfo{year}{2020}).

\bibitem{constraint_graph_coloring}
\bibinfo{author}{Kawakami, S.} \emph{et~al.}
\newblock \bibinfo{title}{A constrained graph coloring solver based on ising machines}.
\newblock In \emph{\bibinfo{booktitle}{2023 IEEE International Conference on Consumer Electronics (ICCE)}}, \bibinfo{pages}{1--6} (\bibinfo{year}{2023}).

\bibitem{inaba2022potts}
\bibinfo{author}{Inaba, K.}, \bibinfo{author}{Inagaki, T.}, \bibinfo{author}{Igarashi, K.} \emph{et~al.}
\newblock \bibinfo{title}{Potts model solver based on hybrid physical and digital architecture}.
\newblock \emph{\bibinfo{journal}{Communications Physics}} \textbf{\bibinfo{volume}{5}}, \bibinfo{pages}{137} (\bibinfo{year}{2022}).

\bibitem{whitehead2023cmos_potts}
\bibinfo{author}{Whitehead, W.}, \bibinfo{author}{Nelson, Z.}, \bibinfo{author}{Camsari, K.} \emph{et~al.}
\newblock \bibinfo{title}{Cmos-compatible ising and potts annealing using single-photon avalanche diodes}.
\newblock \emph{\bibinfo{journal}{Nature Electronics}} \textbf{\bibinfo{volume}{6}}, \bibinfo{pages}{1009--1019} (\bibinfo{year}{2023}).

\bibitem{gc_tabucol}
\bibinfo{author}{Hertz, A.} \& \bibinfo{author}{de~Werra, D.}
\newblock \bibinfo{title}{Using tabu search techniques for graph coloring}.
\newblock \emph{\bibinfo{journal}{Computing}} \textbf{\bibinfo{volume}{39}}, \bibinfo{pages}{345--351} (\bibinfo{year}{1987}).

\bibitem{physics_rev_gc_gnn}
\bibinfo{author}{Schuetz, M. J.~A.}, \bibinfo{author}{Brubaker, J.~K.}, \bibinfo{author}{Zhu, Z.} \& \bibinfo{author}{Katzgraber, H.~G.}
\newblock \bibinfo{title}{Graph coloring with physics-inspired graph neural networks}.
\newblock \emph{\bibinfo{journal}{Phys. Rev. Res.}} \textbf{\bibinfo{volume}{4}}, \bibinfo{pages}{043131} (\bibinfo{year}{2022}).

\bibitem{li2022rethinking_gnn_gc}
\bibinfo{author}{Li, W.} \emph{et~al.}
\newblock \bibinfo{title}{Rethinking graph neural networks for the graph coloring problem}.
\newblock \emph{\bibinfo{journal}{ArXiv}}  (\bibinfo{year}{2022}).
\newblock \eprint{2208.06975}.

\bibitem{koopmans1957assignment_knapsack}
\bibinfo{author}{Koopmans, T.~C.} \& \bibinfo{author}{Beckmann, M.}
\newblock \bibinfo{title}{Assignment problems and the location of economic activities}.
\newblock \emph{\bibinfo{journal}{Econometrica}} \textbf{\bibinfo{volume}{25}}, \bibinfo{pages}{53--76} (\bibinfo{year}{1957}).
\newblock \bibinfo{note}{Accessed 29 Sept. 2024}.

\bibitem{layout_decomposition}
\bibinfo{author}{Kahng, A.~B.}, \bibinfo{author}{Park, C.-H.}, \bibinfo{author}{Xu, X.} \& \bibinfo{author}{Yao, H.}
\newblock \bibinfo{title}{Layout decomposition approaches for double patterning lithography}.
\newblock \emph{\bibinfo{journal}{IEEE Transactions on Computer-Aided Design of Integrated Circuits and Systems}} \textbf{\bibinfo{volume}{29}}, \bibinfo{pages}{939--952} (\bibinfo{year}{2010}).

\bibitem{register_allocation}
\bibinfo{author}{Smith, M.~D.}, \bibinfo{author}{Ramsey, N.} \& \bibinfo{author}{Holloway, G.}
\newblock \bibinfo{title}{A generalized algorithm for graph-coloring register allocation}.
\newblock In \emph{\bibinfo{booktitle}{Proceedings of the ACM SIGPLAN 2004 Conference on Programming Language Design and Implementation}}, PLDI '04, \bibinfo{pages}{277–288} (\bibinfo{publisher}{Association for Computing Machinery}, \bibinfo{address}{New York, NY, USA}, \bibinfo{year}{2004}).

\bibitem{logic_minimization}
\bibinfo{author}{Ciesielski, M.}, \bibinfo{author}{Yang, S.} \& \bibinfo{author}{Perkowski, M.}
\newblock \bibinfo{title}{Multiple-valued boolean minimization based on graph coloring}.
\newblock In \emph{\bibinfo{booktitle}{Proceedings 1989 IEEE International Conference on Computer Design: VLSI in Computers and Processors}}, \bibinfo{pages}{262--265} (\bibinfo{year}{1989}).

\bibitem{earl2005paralleltempering}
\bibinfo{author}{Earl, D.~J.} \& \bibinfo{author}{Deem, M.~W.}
\newblock \bibinfo{title}{Parallel tempering: theory, applications, and new perspectives}.
\newblock \emph{\bibinfo{journal}{Physical Chemistry Chemical Physics}} \textbf{\bibinfo{volume}{7}}, \bibinfo{pages}{3910--3916} (\bibinfo{year}{2005}).

\bibitem{trick2002color_dataset}
\bibinfo{author}{Trick, M.}
\newblock \bibinfo{title}{{COLOR Dataset}} (\bibinfo{year}{2002}).
\newblock \bibinfo{note}{Accessed: 2024-09-29}.

\bibitem{Lewis2021_gc_book}
\bibinfo{author}{Lewis, R. M.~R.}
\newblock \emph{\bibinfo{title}{Guide to Graph Colouring: Algorithms and Applications}}.
\newblock Texts in Computer Science (\bibinfo{publisher}{Springer, Cham}, \bibinfo{year}{2021}), \bibinfo{edition}{2nd} edn.

\bibitem{all_to_kerem_benchmark}
\bibinfo{author}{Nikhar, S.}, \bibinfo{author}{Kannan, S.}, \bibinfo{author}{Aadit, N.~A.}, \bibinfo{author}{Chowdhury, S.} \& \bibinfo{author}{Camsari, K.~Y.}
\newblock \bibinfo{title}{All-to-all reconfigurability with sparse and higher-order ising machines}.
\newblock \emph{\bibinfo{journal}{Nature Communications}} \textbf{\bibinfo{volume}{15}}, \bibinfo{pages}{8977} (\bibinfo{year}{2024}).

\bibitem{sbm_science}
\bibinfo{author}{Goto, H.}, \bibinfo{author}{Tatsumura, K.} \& \bibinfo{author}{Dixon, A.~R.}
\newblock \bibinfo{title}{Combinatorial optimization by simulating adiabatic bifurcations in nonlinear hamiltonian systems}.
\newblock \emph{\bibinfo{journal}{Science Advances}} \textbf{\bibinfo{volume}{5}}, \bibinfo{pages}{eaav2372} (\bibinfo{year}{2019}).
\newblock \eprint{https://www.science.org/doi/pdf/10.1126/sciadv.aav2372}.

\bibitem{sbm_git}
\bibinfo{author}{Ageron, R.}, \bibinfo{author}{Bouquet, T.} \& \bibinfo{author}{Pugliese, L.}
\newblock \bibinfo{title}{Simulated bifurcation (sb) algorithm for python}.
\newblock \bibinfo{howpublished}{\url{https://github.com/bqth29/simulated-bifurcation-algorithm}} (\bibinfo{year}{2023}).
\newblock \bibinfo{note}{Version 1.2.1, Nov. 2023}.

\bibitem{dsatur}
\bibinfo{author}{Gualandi, S.} \& \bibinfo{author}{Malucelli, F.}
\newblock \bibinfo{title}{Exact solution of graph coloring problems via constraint programming and column generation}.
\newblock \emph{\bibinfo{journal}{INFORMS Journal on Computing}} \textbf{\bibinfo{volume}{24}}, \bibinfo{pages}{81--100} (\bibinfo{year}{2012}).
\newblock \eprint{https://doi.org/10.1287/ijoc.1100.0436}.

\bibitem{Hamerly2019ExperimentalAnnealer}
\bibinfo{author}{Hamerly, R.} \emph{et~al.}
\newblock \bibinfo{title}{{Experimental investigation of performance differences between coherent Ising machines and a quantum annealer}}.
\newblock \emph{\bibinfo{journal}{Science Advances}} \textbf{\bibinfo{volume}{5}}, \bibinfo{pages}{eaau0823} (\bibinfo{year}{2019}).

\bibitem{fpga_sbm}
\bibinfo{author}{Tatsumura, K.}, \bibinfo{author}{Dixon, A.~R.} \& \bibinfo{author}{Goto, H.}
\newblock \bibinfo{title}{Fpga-based simulated bifurcation machine}.
\newblock In \emph{\bibinfo{booktitle}{2019 29th International Conference on Field Programmable Logic and Applications (FPL)}}, \bibinfo{pages}{59--66} (\bibinfo{year}{2019}).

\bibitem{git_data}
\bibinfo{author}{Garg, C.} \& \bibinfo{author}{Salahuddin, S.}
\newblock \bibinfo{title}{Optimization\_integer\_ising\_data} (\bibinfo{year}{2025}).
\newblock \bibinfo{note}{\url{https://doi.org/10.5281/zenodo.16945085}}.

\bibitem{ackley1985learning}
\bibinfo{author}{Ackley, D.~H.}, \bibinfo{author}{Hinton, G.~E.} \& \bibinfo{author}{Sejnowski, T.~J.}
\newblock \bibinfo{title}{A learning algorithm for boltzmann machines}.
\newblock \emph{\bibinfo{journal}{Cognitive Science}} \textbf{\bibinfo{volume}{9}}, \bibinfo{pages}{147--169} (\bibinfo{year}{1985}).

\bibitem{bremaud1999gibbs}
\bibinfo{author}{Br\'{e}maud, P.}
\newblock \bibinfo{title}{Gibbs fields and monte carlo simulation}.
\newblock In \emph{\bibinfo{booktitle}{Markov Chains}}, \bibinfo{pages}{253--322} (\bibinfo{publisher}{Springer New York}, \bibinfo{address}{New York, NY}, \bibinfo{year}{1999}).

\bibitem{tsp_comm_si2024energy}
\bibinfo{author}{Si, e.~a.}
\newblock \bibinfo{title}{Energy-efficient superparamagnetic ising machine and its application to traveling salesman problems}.
\newblock \emph{\bibinfo{journal}{Nature Communications}} \textbf{\bibinfo{volume}{15}}, \bibinfo{pages}{3016} (\bibinfo{year}{2024}).

\bibitem{cora}
\bibinfo{author}{McCallum, A.~K.}, \bibinfo{author}{Nigam, K.}, \bibinfo{author}{Rennie, J.} \& \bibinfo{author}{Seymore, K.}
\newblock \bibinfo{title}{Automating the construction of internet portals with machine learning}.
\newblock \emph{\bibinfo{journal}{Information Retrieval}} \textbf{\bibinfo{volume}{3}}, \bibinfo{pages}{127--163} (\bibinfo{year}{2000}).

\bibitem{citseer}
\bibinfo{author}{Sen, P.} \emph{et~al.}
\newblock \bibinfo{title}{Collective classification in network data}.
\newblock \emph{\bibinfo{journal}{AI Magazine}} \textbf{\bibinfo{volume}{29}}, \bibinfo{pages}{93--106} (\bibinfo{year}{2008}).

\bibitem{pubmed}
\bibinfo{author}{Namata, G.}, \bibinfo{author}{London, B.}, \bibinfo{author}{Getoor, L.} \& \bibinfo{author}{Huang, B.}
\newblock \bibinfo{title}{Query-driven active surveying for collective classification}.
\newblock In \emph{\bibinfo{booktitle}{Proceedings of the 10th International Workshop on Mining and Learning with Graphs}}, vol.~\bibinfo{volume}{8}, \bibinfo{pages}{249--256} (\bibinfo{year}{2012}).

\bibitem{pcircuits_datta}
\bibinfo{author}{Li, M.-C.} \emph{et~al.}
\newblock \bibinfo{title}{12.2 p-circuits: Neither digital nor analog}.
\newblock In \emph{\bibinfo{booktitle}{2025 IEEE International Solid-State Circuits Conference (ISSCC)}}, vol.~\bibinfo{volume}{68}, \bibinfo{pages}{1--3} (\bibinfo{year}{2025}).

\bibitem{lawler1986tsp}
\bibinfo{author}{Lawler, E.~L.}
\newblock \bibinfo{title}{The traveling salesman problem: A guided tour of combinatorial optimization}.
\newblock \emph{\bibinfo{journal}{Journal of the Operational Research Society}} \textbf{\bibinfo{volume}{37}}, \bibinfo{pages}{535--536} (\bibinfo{year}{1986}).

\bibitem{pnas_dna}
\bibinfo{author}{Pevzner, P.~A.}, \bibinfo{author}{Tang, H.} \& \bibinfo{author}{Waterman, M.~S.}
\newblock \bibinfo{title}{An eulerian path approach to dna fragment assembly}.
\newblock \emph{\bibinfo{journal}{Proceedings of the National Academy of Sciences}} \textbf{\bibinfo{volume}{98}}, \bibinfo{pages}{9748--9753} (\bibinfo{year}{2001}).
\newblock \eprint{https://www.pnas.org/doi/pdf/10.1073/pnas.171285098}.

\bibitem{dac_tsp}
\bibinfo{author}{Dan, A.}, \bibinfo{author}{Shimizu, R.}, \bibinfo{author}{Nishikawa, T.}, \bibinfo{author}{Bian, S.} \& \bibinfo{author}{Sato, T.}
\newblock \bibinfo{title}{Clustering approach for solving traveling salesman problems via ising model based solver}.
\newblock In \emph{\bibinfo{booktitle}{2020 57th ACM/IEEE Design Automation Conference (DAC)}}, \bibinfo{pages}{1--6} (\bibinfo{year}{2020}).

\bibitem{reinelt1991tsplib}
\bibinfo{author}{Reinelt, G.}
\newblock \bibinfo{title}{Tsplib—a traveling salesman problem library}.
\newblock \emph{\bibinfo{journal}{INFORMS Journal on Computing}} \textbf{\bibinfo{volume}{3}}, \bibinfo{pages}{376--384} (\bibinfo{year}{1991}).

\bibitem{lin1973heuristic}
\bibinfo{author}{Lin, S.} \& \bibinfo{author}{Kernighan, B.~W.}
\newblock \bibinfo{title}{An effective heuristic algorithm for the traveling salesman problem}.
\newblock \emph{\bibinfo{journal}{Operations Research}} \textbf{\bibinfo{volume}{21}}, \bibinfo{pages}{498--516} (\bibinfo{year}{1973}).

\end{thebibliography}


\begin{addendum}

\item [Acknowledgements] 
This work is supported by the Office of Naval Research (ONR), Multidisciplinary University Research Initiative (MURI) grant N000142312708.

\item [Author Contributions] 
C.G. formulated the vectorized scheme, performed the CPU/GPU benchmarks and designed the vectorized accelerator on FPGA. C.G, and S.S co-wrote the manuscript; S.S supervised the research. 
All authors contributed to discussions and commented on the manuscript. 

\item [Competing Interests] 
The authors declare that they have no competing financial interests.

\end{addendum}
\clearpage


\begin{algorithm}
\caption{Vectorized Mapping for Graph Coloring Problem ($F$-operator):}
\begin{algorithmic}[1]
\label{Algorithm_1}
\STATE $Q \gets \text{number of colors}$
\STATE $H \gets \text{Hamiltonian Energy}$
\STATE $W_{ij} \gets \text{connectivity weight between node i and j}$
\STATE $n \gets \left\lceil \log_2 (Q) \right\rceil $
\STATE $S_i \gets \text{color vector for node i }  \{  s_{i0}, s_{i1}, \ldots, s_{i(n-1)} \}$\\
\STATE Generate $F$ operator in truth table format for any two graph nodes ($S_i$ $S_j$):
    \STATE $ \text{select variables} \gets \{  s_{i0}, s_{i1}, \ldots, s_{i(n-1)},  s_{j0}, s_{j1}, \ldots, s_{j(n-1)} \} $
    \IF{$S_i == S_j$}
        \STATE $F \gets 1 \text{ [two nodes colored with same color]}$ 
        \STATE $H \gets H + W_{ij}$
    \ELSIF{$S_i, S_j \notin [0, Q - 1]$}
        \STATE $F \gets 1 \text{ [assigned color out of range]}$ 
        \STATE $H \gets H + W_{ij}$
    \ELSE
        \STATE $F \gets 0 \text{ [two nodes colored with different color]}$
        \STATE $H \gets H +0$
    \ENDIF
    
\end{algorithmic}
\end{algorithm}

\clearpage

\begin{algorithm}
\caption{Probabilistic Ising Machine Spin Update}
\begin{algorithmic}[1]
\label{Algorithm_2}
\STATE $H \gets \text{Hamiltonian} $
\STATE $T \gets \text{temperature}$
\STATE $\beta \gets {1}/{T}$
\STATE $N_S \gets \text{Number of iteration (update) steps} $
\STATE $k \gets \text{Total Spins}$
\STATE $sigmoid(x) = 1/(1+e^{-x})$
\STATE Initialize the spins s $\{  s_{0}, s_{1}, \ldots, s_{k}\}$ $\in$ [0, 1] randomly
\FOR{$t=0$ to $t = N_S$}
    \STATE $i \gets 1$
    \FOR{$i=0$ to $t = k$}
        \STATE Calculate $\Delta H_i = H_{s_i=1}- H_{s_i=0}$
        \STATE $r \gets \text{random number}$
        \IF{$sigmoid(-\Delta H_{i}/T) > r$}
            \STATE $s_i \gets 1 $
        \ELSE
            \STATE $s_i \gets 0 $
        \ENDIF
    \ENDFOR
\ENDFOR

\end{algorithmic}
\end{algorithm}
\clearpage



\begin{table}[ht]
\centering
\caption{Solution accuracy in terms of wrongly colored edges (lower value is better) comparison of heuristic (Tabucol), learning-based approaches (GNN and PI-SAGE), Ising methods, and Vectorized framework. NA is mentioned for problem instances not reported in \cite{physics_rev_gc_gnn}.}
\label{fig:accuracy_table}
\begin{adjustbox}{width=\textwidth}


\centering
\rowcolors{2}{cyan!5}{white}  

\begin{tabular}{|c|c|c|c||c|c||c||c|c|c|c||c|c|}
\hline
\hline

\textbf{Problem} & \textbf{$\#$nodes} & \textbf{$\#$edges} & \textbf{$\#$colors}  & \textbf{$\#$GNN\cite{physics_rev_gc_gnn}} & \textbf{$\#$PI-SAGE\cite{physics_rev_gc_gnn}} & \textbf{$\#$Tabucol} & \multicolumn{1}{|c|}{\shortstack{\textbf{$\#$Probabilistic} \\ \textbf{Ising}}} & \multicolumn{1}{|c|}{\shortstack{\textbf{$\#$Simulated} \\ \textbf{Bifurcation}}} &
\multicolumn{1}{|c|}{\shortstack{\textbf{$\#$Vectorized} \\ \textbf{GPU}}} & \multicolumn{1}{|c||}{\shortstack{\textbf{$\#$Vectorized} \\ \textbf{FPGA}}} &
\multicolumn{1}{|c|}{\shortstack{\textbf{$\#$ Probabilistic Ising + } \\ \textbf{Parallel Tempering GPU}}} &
\multicolumn{1}{|c|}{\shortstack{\textbf{$\#$Vectorized+ Parallel} \\ \textbf{Tempering GPU}}}
\\
\hline
\hline
 anna & 138 & 493 & 11 & 1 & 0 & 0 & 12 & 44 & 0 & 0 & 2 & \textbf{0}\\
\hline
david & 87 & 406 & 11 & NA & NA & 0 & 17 & 11 & 0 & 0 & 10 & \textbf{0} \\
\hline
huck & 74 & 301 & 11 & NA & NA & 0 & 0 & 13 & 0 & 0 & 0 &  \textbf{0} \\
\hline
myciel3 & 11 & 20 & 4 & NA & NA & 0 & 0 & 0 & 0 & 0 & 0 &  \textbf{0} \\
\hline
myciel4 & 23 & 71 & 5 & NA & NA & 0 & 1 & 0 & 0 & 0 & 0 &  \textbf{0} \\
\hline
myciel5 & 47 & 236 & 6 & 0 & 0 & 0 & 0 & 0 & 0 & 0 & 0 &  \textbf{0} \\
\hline
myciel6 & 95 & 755 & 7 & 0 & 0 & 0 & 4 & 6 & 0 & 0 & 2 &  \textbf{0} \\
\hline
myciel7 & 191 & 2360 & 8 & NA & NA & 0 & 144 & 61 & 0 & 0 & 52 &  \textbf{0} \\
\hline
queen5$\_$5 & 25 & 160 & 5 & 0 & 0 & 0 & 5 & 5 & 0 & 0 & 0 &  \textbf{0} \\
\hline
queen6$\_$6 & 36 & 290 & 7 & 4 & 0 & 0 & 3 & 6 & 1 & 1 & 1 &  \textbf{0} \\
\hline
queen7$\_$7 & 49 & 476 & 7 & 15 & 0 & 0 & 21 & 24 & 6 & 5 & 14 & \textbf{0} \\
\hline
queen8$\_$8 & 64 & 728 & 9 & 7 & 1 & 0 & 41 & 29 & 4 & 2 & 15 & \textbf{1} \\
\hline
queen9$\_$9 & 81 & 1056 & 10 & 13 & 1 & 0 & 36 & 47 & 5 & 4 & 18 & \textbf{2} \\
\hline
queen8$\_$12 & 96 & 1368 & 12 & 7 & 0 & 0 & 44 & 73 & 2 & 2 & 21 & \textbf{0} \\
\hline
queen11$\_$11 & 121 & 1980 & 11 & 33 & 17 & 15 & 87 & 111 & 20 & 18 & 41 & \textbf{14} \\
\hline
queen13$\_$13 & 169 & 3328 & 13 & 40 & 26 & 21 & 124 & 199 & 31 & 26 & 60 & \textbf{21} \\
\hline
\hline

\end{tabular}

\end{adjustbox}
\end{table}

\clearpage

\begin{table}[ht]
\centering
\caption{Solution accuracy for citation graphs in terms of wrongly colored edges (lower value is better) comparison of heuristic (Tabucol), learning-based approaches (GNN and PI-SAGE), Ising methods, and Vectorized framework. NA is mentioned for a problem instance that takes more than 24 hrs to run or when the GPU runs out of resources. }
\label{fig:accuracy_table_citation}
\begin{adjustbox}{width=\textwidth}


\centering
\rowcolors{2}{cyan!5}{white}  

\begin{tabular}{|c|c|c|c||c|c||c||c|c|c|c|c||c|c|}
\hline
\hline

\textbf{Problem} & \textbf{$\#$nodes} & \textbf{$\#$edges} & \textbf{$\#$colors}  & \textbf{$\#$GNN\cite{physics_rev_gc_gnn}} & \textbf{$\#$PI-SAGE\cite{physics_rev_gc_gnn}} & \textbf{$\#$Tabucol} & \multicolumn{1}{|c|}{\shortstack{\textbf{$\#$Probabilistic} \\ \textbf{Ising}}} &  \multicolumn{1}{|c|}{\shortstack{\textbf{$\#$Simulated} \\ \textbf{Bifurcation}}} &
\multicolumn{1}{|c|}{\shortstack{\textbf{$\#$Vectorized} \\ \textbf{GPU}}} & 
\multicolumn{1}{|c|}{\shortstack{\textbf{$\#$ Probabilistic Ising + } \\ \textbf{Parallel Tempering GPU}}} &
\multicolumn{1}{|c|}{\shortstack{\textbf{$\#$Vectorized+ Parallel} \\ \textbf{Tempering GPU}}}
\\
\hline
\hline
cora & 2708 & 5278 & 5 & 3 & 0 & 0 & 1004 & 486 & 2 & 830 & 1\\
\hline
citseer & 3279 & 4552 & 6 & 3 & 0 & 0  & 707 & 163 & 1 & 592 & 0 \\
\hline
pubmed & 19717 & 44324 & 8 & 35 & 17 & 17 & NA & NA & 18 & NA & 11  \\
\hline
\hline
\hline

\end{tabular}

\end{adjustbox}
\end{table}
\clearpage


\clearpage

\clearpage

\clearpage

\begin{figure*}
\begin{centering}
\includegraphics[width=1\linewidth]{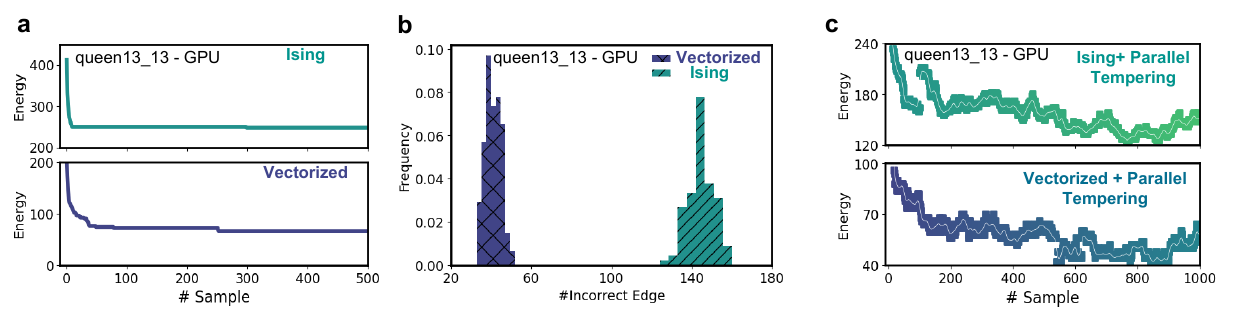}
\par\end{centering}
\caption{\label{fig:solution_exploration} \textbf{Solution Exploration in Ising and Vectorized Mapping framework.} \\
\textbf{a,} Energy evolution for $queen13\_13$ problem instance mapped and solved using Ising and Vectorized mapping implementation on GPU.
{\textbf{(b)}} Distribution of number of incorrectly colored edges achieved for $queen13\_13$ problem instance after running GPU-based Ising and Vectorized implementation for 200 parallel runs.
{\textbf{(c)}} Ising and Vectorized framework combined with parallel tempering enhances the solution exploration avoiding the states being stuck around local minima solution. 
}

\end{figure*}

\clearpage

\begin{figure*}
\begin{centering}
\includegraphics[width=\linewidth]{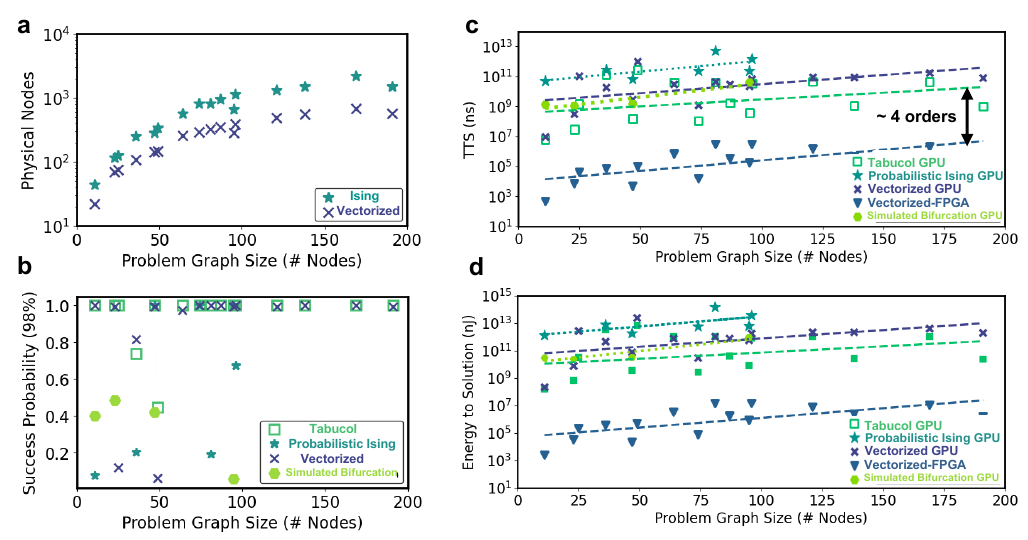}
\par\end{centering}
\caption{\label{fig:area_TTS} \textbf{Area, Solution Quality, Performance and Energy Efficiency Benchmarks} 
\\
\textbf{a,} Physical implementation nodes used in Ising and Vectorized mapping approach for dataset problem \cite{trick2002color_dataset}. 
 {\textbf{(b)}} Success Probability (> 98 $\%$ correctly colored edges) metric comparing vectorized mapping with Tabucol heuristics and and physics-based Ising solvers including Probabilistic Ising and Simulated Bifurcation Machines.
 {\textbf{(c)}} Time-to-solution (TTS) and {\textbf{(d)}} Energy-to-solution (ETS) benchmark of vectorized mapping implemented on FPGA with Tabucol heuristic and Ising solvers implementation on GPU.
}

\end{figure*}





\onecolumn
\clearpage

\beginsupplement

\begin{center}
\LARGE \textbf{Supplementary Information} \\[1.5em]
\LARGE \textbf{Efficient Optimization Accelerator Framework for Multi-state Spin Ising Problems} \\[1.5em]

\normalsize
Chirag Garg$^{1}$ \faEnvelope[regular], Sayeef Salahuddin$^{1}$ \text{\faEnvelope[regular]} \\[1.5em]

$^{1}$Department of Electrical Engineering and Computer Sciences, University of California, Berkeley, CA 94720, USA. \\[0.8em]

\textbf{* Corresponding author. Email: chirag$\_$garg@berkeley.edu; sayeef@berkeley.edu}

\end{center}

\clearpage

\section*{1. Probabilistic Ising Machines}

Probabilistic Ising machines follow the principle of the Boltzmann Machine binary neural network \cite{ackley1985learning}. Therefore, the probability distribution corresponding to a given state $p(s) = \frac{1}{Z} e^{-H(s)/T}$ where $Z = \sum_{s}e^{-H(s)/T}$ is the normalizing/partition function, $s$ represents the states, T denotes temperature coefficient and $H(s)$ is Hamiltonian or energy function to be minimized. Ising Hamiltonian is generally of the form $H(s) = \sum_i \sum_j W_{ij}s_is_{j} + \sum_i b_i s_i$, where $s_i$ and $s_{j}$ represents the state of spins taking binary values \{0, 1\}, $W_{ij}$ sets the interaction weight between these spins and $b_{i}$ is bias value for spin $s_i$. The probability distribution represented by this Ising Hamiltonian is sampled using Gibbs sampling \cite{bremaud1999gibbs} such that the Ising machine stochastically moves toward higher probability states and lower energy. It is enabled by appropriately formulating the update rule, which for Ising Hamiltonian, is given by $P(s_i = 1| s) = sigmoid (- \Delta H_{i}/T)$ where $sigmoid(x) = 1/(1+e^{-x})$. 

This work particularly employs single flip Gibbs sampling to update the spins as illustrated in Algorithm \ref{Algorithm_2}. For each spin update, the algorithm first calculates the change in Hamiltonian ($\Delta H$) due to spin flip and compares $sigmoid (- \Delta H_{i}/T)$ with a random number from uniform distribution of [0, 1] to update the spin. This comparison predominantly supports the spin flips that cause Hamiltonian to minimize. However, there is still a non-zero probability that it would accept spin updates that cause the Hamiltonian to increase \cite{ackley1985learning, tsp_comm_si2024energy, supriyo2019integer}. It gives rise to minor energy fluctuations in energy exploration data reported in Supplementary Fig. \ref{fig:energy_evolution}.

\section*{2. Hyperparameter optimization for graph coloring}

\begin{figure*}[h!]
\begin{centering}
\includegraphics[width=\linewidth]{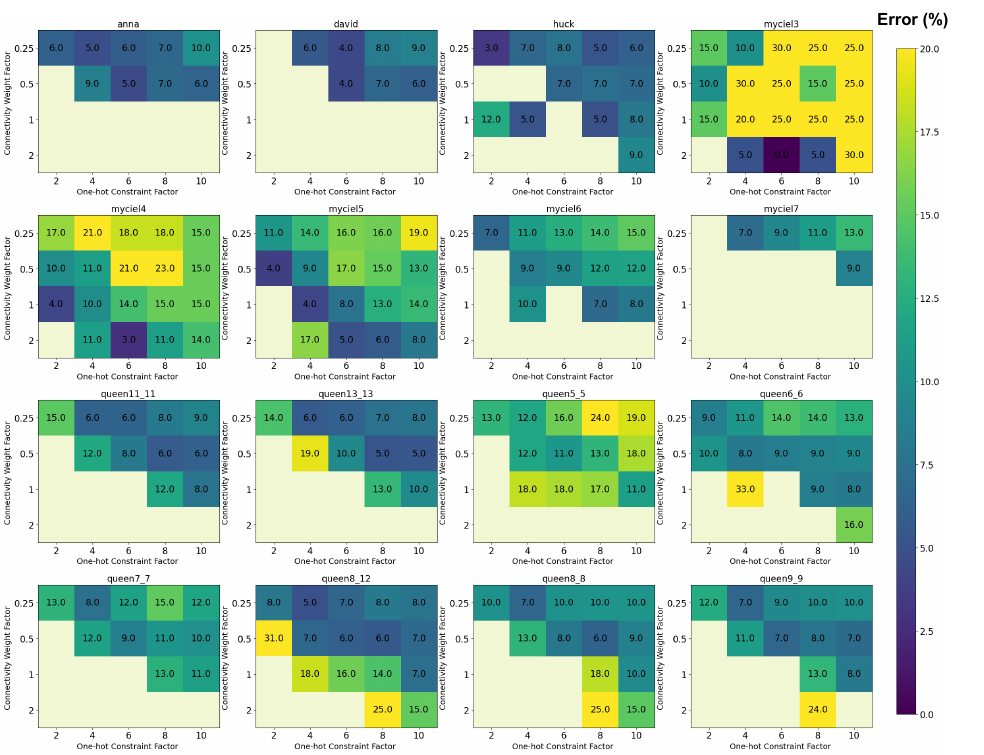}
\par\end{centering}
\caption{\label{fig:hyperparam} Parameter Search across graph coloring problem instances for probabilistic Ising framework. It searches for the optimal value of the connectivity weight factor and one-hot constraint factor to minimize the error. Error is defined as incorrectly colored edges divided by total edges in the problem graph.
}
\end{figure*}

In order to make a fair comparison between Ising and vectorized mapping, a grid search-based parameter optimization is performed for the parameters $A$ in Eq. \ref{eq:graph_coloring} and $B$ in Eq. \ref{eq:coloring_constraint} in the graph coloring Hamiltonian. If the connectivity weighting factor ($A$) for this constraint is set relatively higher, the Ising machine may frequently violate the one-hot encoding condition. This can lead to the exploration of invalid solution spaces or failure to enforce correct color assignments, resulting in incorrect graph colorings. On the other hand, if one-hot constraint factor ($B$) is too high, the machine may place too much emphasis on satisfying the one-hot constraint, which can hinder the optimization of the actual coloring objective. As a result, it may become trapped in poor local minima where valid but suboptimal solutions prevail. Supplementary Fig. \ref{fig:hyperparam} reports the hyperparameter optimization where we report parameter search for each problems separately by running it for 1000 iteration steps at $T$ = 0.2 and reporting the coloring error (incorrectly colored edges divided by total edges). The optimal settings from Supplementary Fig. \ref{fig:hyperparam} are chosen in this work for testing graph coloring problem instances using the probabilistic Ising framework. 
The simulated bifurcation framework\cite{sbm_git} used in this work only accepts the Ising Hamiltonian to minimize. Accordingly, the parameters $A$ in Eq. \ref{eq:graph_coloring} and $B$ in Eq. \ref{eq:coloring_constraint} are optimized by running the framework for 10000 iteration steps. The optimal values identified in Supplementary Fig. \ref{fig:SBM_hyperparam} are then used to construct the Hamiltonian, which is subsequently solved using the simulated bifurcation method for the graph coloring problem. For vectorized mapping implementation, only temperature $T$ effects the results and is chosen equal to 0.2 for all problem instances. 

\section*{3. Vectorized Mapping Benchmark for Citation Graph Dataset}
\begin{figure*}[h!]
\begin{centering}
\includegraphics[width=0.5\linewidth]{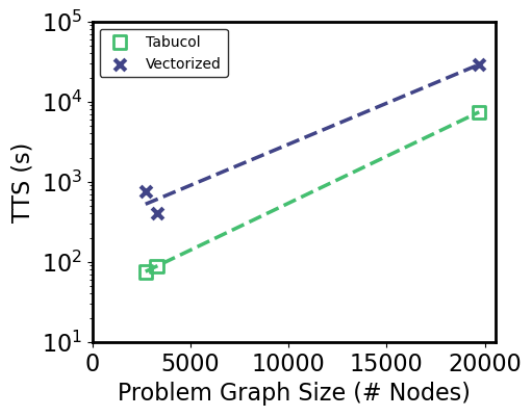}
\par\end{centering}
\caption{\label{fig:tts_citation} Time-to-solution (TTS) for citation graphs
benchmarks of vectorized mapping with Tabucol heuristic both implementated on GPU.
}
\end{figure*}

To demonstrate the scalability of vectorized mapping approach, we benchmark the citation datasets (Cora \cite{cora}, Citeseer \cite{citseer}, and Pubmed \cite{pubmed}). These problems are often used for graph-based benchmark experiments and have also been used for testing graph-coloring based algorithms \cite{physics_rev_gc_gnn, li2022rethinking_gnn_gc}. Table \ref{fig:accuracy_table_citation} confirms that the solution accuracy advantage of the proposed vectorized mapping holds for large size graph problems. Supplementary Fig. \ref{fig:tts_citation} shows the time-to-solution for these citation graphs. It confirms that the vectorized mapping on GPU takes around one order of magnitude more time compared to the Tabucol heuristics aligning with the scaling trends in Fig. \ref{fig:area_TTS}c. Moreover, in probabilistic Ising hardware, the overall hardware scaling is primarily influenced by the number of physical nodes required to solve a problem \cite{pcircuits_datta, saavan2024pass}. As a result, it follows the scaling behavior of physical nodes with respect to problem size illustrated in Fig. \ref{fig:area_TTS}a.

\section*{4. Traveling Salesman Problem}

The Traveling Salesman Problem (TSP) is a classic NP-hard optimization problem that aims to determine the shortest possible route in which a given set of cities is visited exactly once. Owing to its computational complexity and broad applicability in areas such as logistics \cite{lawler1986tsp}, circuit design\cite{simulated_annealing_kirkpatrick}, and DNA sequencing\cite{pnas_dna}, TSP remains a central focus in combinatorial optimization research. Recently, Ising machines are being explored to tackle this problem \cite{whitehead2023cmos_potts, dac_tsp}. To solve $N$-cities problems, these solvers require $N^2$ spins modeled via a lattice-like graph representing city number and spin. The Hamiltonian is given as follows: 
\begin{equation}
\renewcommand{\theequation}{S1}
    \label{eq:tsp_hamiltonian}
    H = A\sum_{k \neq l }\sum_{i}W_{kl}s_{ik}s_{(i+1)l}+ B\sum_{i}
    (1-\sum_{k}s_{ik})^{2} + B\sum_{k}
    (1-\sum_{i}s_{ik})^{2} \\ 
\end{equation}
where $A$ is the connectivity weight factor, $B$ one-hot constraint factor, $W_{kl}$ represents weight defined as distance between city k and l, and $s_{ik}$ $\in \{0, 1\}$ is the spin value for visiting $k$ city at $i^{th}$ position in the tour. The first term in the Hamiltonian represents the total weight for given values of spins. However, the second and third terms are to enforce one-hot encoding constraints that penalize assigning multiple cities for visiting at the same  $i^{th}$ position and visiting a city multiple times in a tour, respectively. In this work, we use the $burma14$ problem instance from the tsplib dataset \cite{reinelt1991tsplib} and use the first subset of the weight matrix to generate smaller problem instances. The weight matrix is normalized with the maximum distance in it \cite{dac_tsp}. For the Ising framework, we employ Gibbs Sampling described in Algorithm \ref{Algorithm_2} and choose temperature value 0.02. Further, we exploit the grid-search hyperparameter optimization for parameters $A$ and $B$ as shown in Supplementary Fig. \ref{fig:hyperparam_tsp}. Each problem instance is run for 4000 iteration steps, and optimized for the optimality gap. It is given by the Supplementary Eq. \ref{eq:optimality_gap}, where the optimal solution is calculated using the Lin-Kernighan (LK) heuristics \cite{lin1973heuristic}. 
\begingroup

\renewcommand{\theequation}{S2}

\begin{equation}
    \label{eq:optimality_gap}
    \text{Optimality Gap} = 1 - \frac{\text{tour cost obtained by algorithm}}{\text{optimal tour cost } } \\ 
\end{equation}
\endgroup

We propose the vectorized mapping framework to tackle the TSP problem, which maps $N$ city problem to $\left\lceil Nlog_{2}(N)\right\rceil $. A spin vector $S_i$  represents the position of the city $i$ in the tour and is defined as $\{s_{i0}, s_{i1}.. s_{i(n-1)}\}$ where  $n = \left\lceil log_{2}(N)\right\rceil$. Supplementary Algorithm \ref{Algorithm_3} describes the formulation of vectorized mapping Hamiltonian and operator function $F$. If $S_i$ and $S_j$ are consecutive cities, the Hamiltonian is incentivized to optimize for the tour cost. However, it is penalized when two cities are visited at the same position in the tour guided by the hyperparameter $wt$. Further, the formulation also restricts $S_i$ to be within its range of N. 
For the vectorized mapping framework, we employ the Gibbs sampling described in Algorithm \ref{Algorithm_2} and choose the temperature value 0.2. The hyperparameter $wt$ is optimized for the optimality gap, as shown in Supplementary Fig. \ref{fig:hyperparam_tsp_vec}.

To evaluate the broader applicability of the proposed vectorized mapping framework, we conducted additional experiments on the Traveling Salesman Problem (TSP). Both the Ising model and the proposed vectorized mapping were implemented on Nvidia A100 Tensor Core GPU and tested across TSP instances involving up to 14 cities. Each instance was solved 500 times, involving 4000 iteration or update steps per run. As shown in Supplementary Fig. \ref{fig:tsp_ising_vec}, the proposed approach achieves higher success probabilities and shorter time-to-solution compared to the Ising model. Additionally, it consistently produces tours that are closer to those found by the well-established Lin–Kernighan (LK) heuristic, indicating improved convergence toward high-quality solutions (see Supplementary Fig. \ref{fig:optimality_gap}). These results suggest that the vectorized mapping framework is an effective strategy for addressing complex combinatorial optimization problems beyond graph coloring, such as TSP.

\clearpage
\begin{algorithm}
\renewcommand{\thealgorithm}{A1}
\caption{Vectorized Mapping for Traveling Salesman Problem ($F$-operator):}
\begin{algorithmic}[1]
\label{Algorithm_3}
\STATE $N \gets \text{total cities}$
\STATE $H \gets \text{Energy Hamiltonian}$
\STATE $wt \gets \text{hyperparameter}$
\STATE $n \gets \left\lceil \log_2 (N) \right\rceil $
\STATE $W \gets \text{weight matrix for traveling salesman problem}$
\STATE $W_{ij} \gets \text{normalized distance between city i and j}$
\STATE $S_i \gets \text{position vector for city $i$ in the tour}  \{  s_{i0}, s_{i1}, \ldots, s_{i(n-1)} \}$\\
\STATE Generate $F$ operator in truth table format for any two city nodes ($S_i$ $S_j$):
    \STATE $ \text{input variables} \gets \{  s_{i0}, s_{i1}, \ldots, s_{i(n-1)},  s_{j0}, s_{j1}, \ldots, s_{j(n-1)} \} $
    \IF{$|S_i - S_j|==1$}
        \STATE $F \gets 1 \text{ [$S_i$ and $S_j$ cities are next to each other in tour]}$ 
        \STATE $H \gets H + W_{ij}$
    \ELSIF{$S_i== S_j$}
        \STATE $F \gets -1 \text{ [$S_i$ and $S_j$ cities are assigned same position in the tour]}$ 
        \STATE $H \gets H + wt*(\sum_{i} W_{ij}) \text{ [penalize the energy hamiltonian]}$  
    \ELSIF{$(S_i,  S_j) \notin [0, N-1]$}
        \STATE $F \gets -2 \text{ [$S_i$ and $S_j$ cities are assigned outside the time vector range]}$ 
        \STATE $H \gets H + max (W_{ij}) \text{ [penalize the energy hamiltonian]}$  
    \ELSE
        \STATE $F \gets 0 \text{ [two cites are not next to each other in the tour]}$ 
        \STATE $H \gets 0 \text{ [does not affect the energy hamiltonian]}$ 
    \ENDIF
    
\end{algorithmic}
\end{algorithm}


\begin{figure*}[h!]
\begin{centering}
\includegraphics[width=\linewidth]{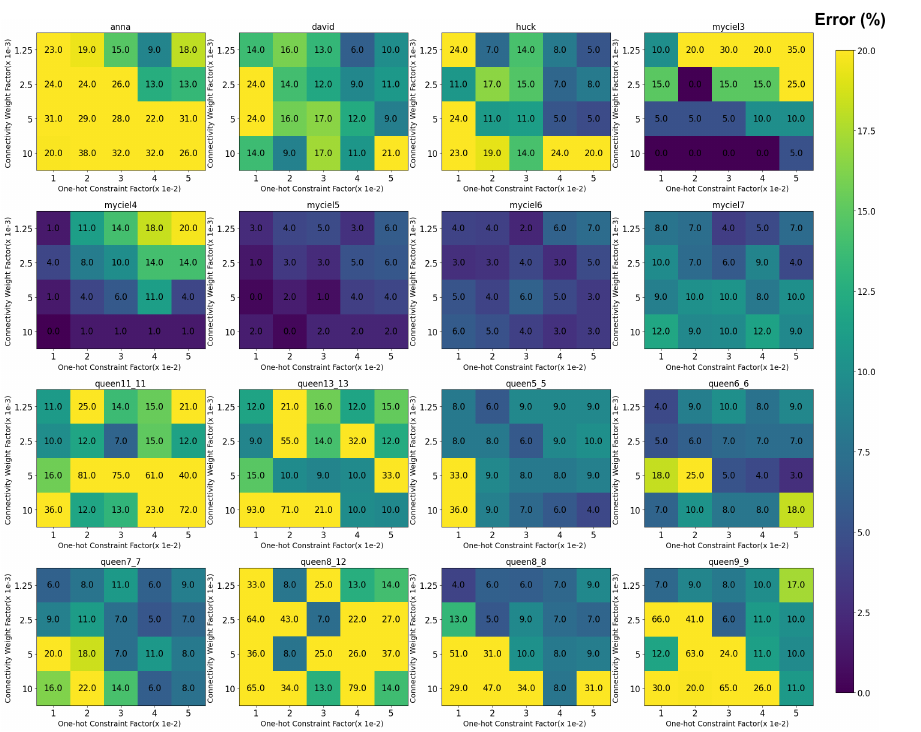}
\par\end{centering}
\caption{\label{fig:SBM_hyperparam} Parameter Search across graph coloring problem instances for Simulated Bifurcation framework. It searches for the optimal value of the connectivity weight factor and one-hot constraint factor to minimize the error. Error is defined as incorrectly colored edges divided by total edges in the problem graph.
}
\end{figure*}


\begin{figure*}[h!]
\begin{centering}
\includegraphics[width=\linewidth]{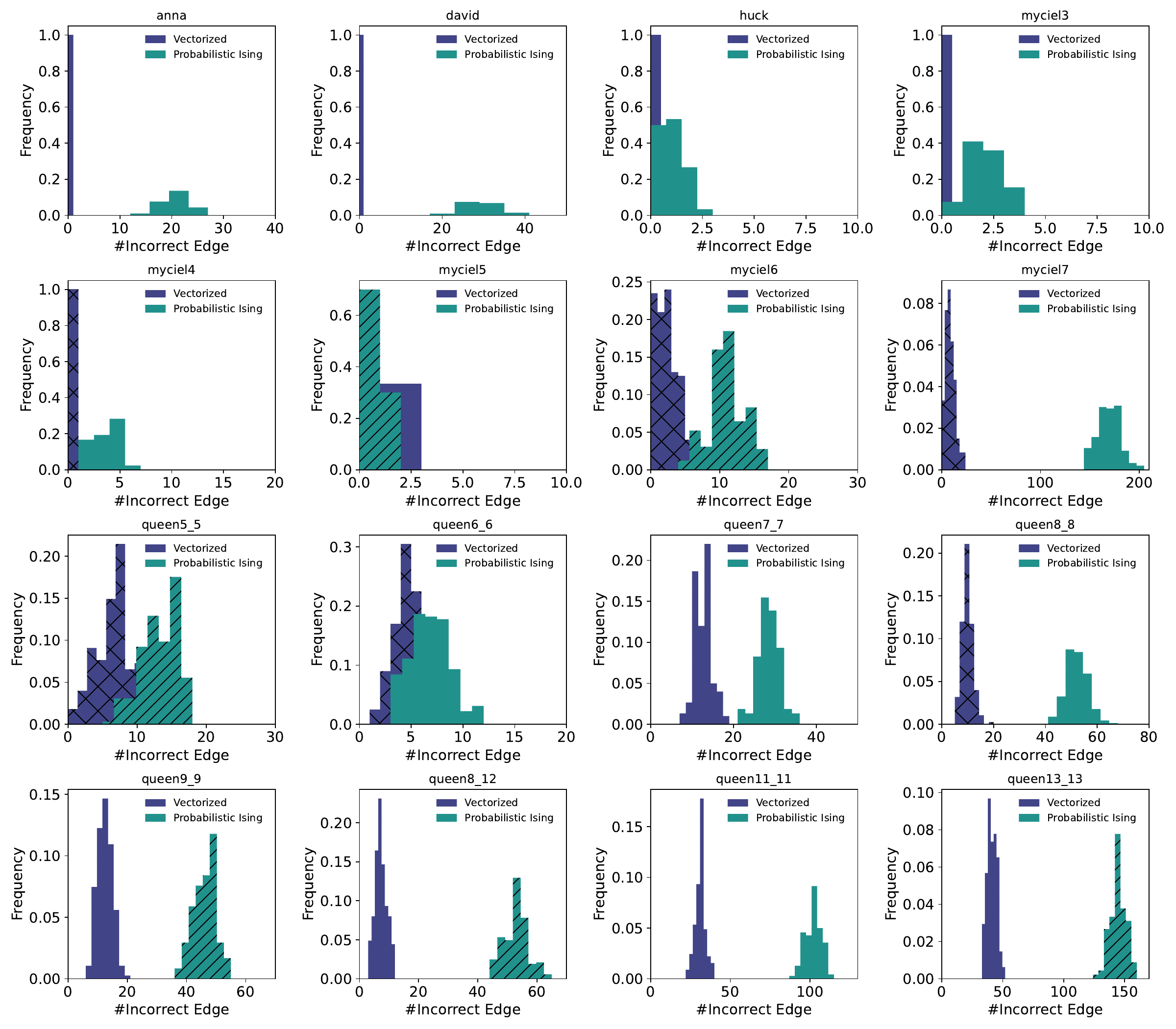}
\par\end{centering}
\caption{\label{fig:incorr_edge} Distribution of number of incorrectly colored edges achieved after completing each of 200 parallel runs while solving the graph coloring problem instances \cite{trick2002color_dataset} on Ising and Vectorized mapping framework.
}
\end{figure*}


\begin{figure*}[h!]
\begin{centering}
\includegraphics[width=\linewidth]{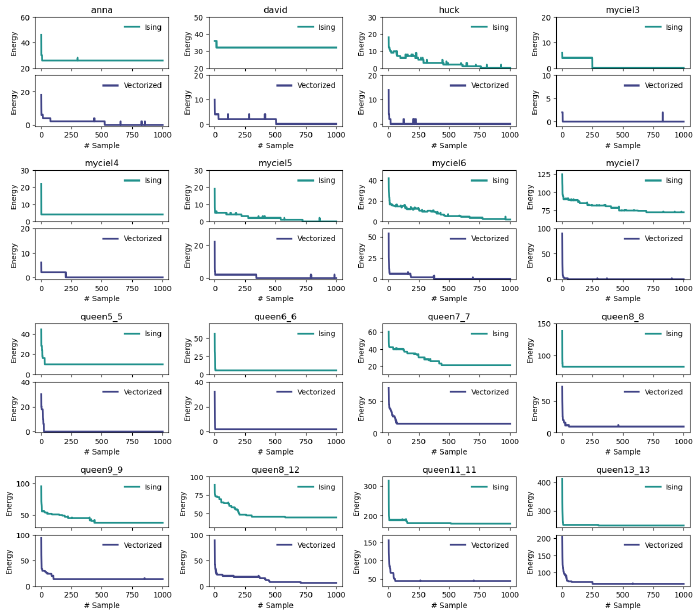}
\par\end{centering}
\caption{\label{fig:energy_evolution} Evolution of energy while solving the graph coloring problem instances \cite{trick2002color_dataset} on Probabilistic Ising and Vectorized mapping framework.
}
\end{figure*}

\clearpage

\clearpage


\begin{figure*}[h!]
\begin{centering}
\includegraphics[width=\linewidth]{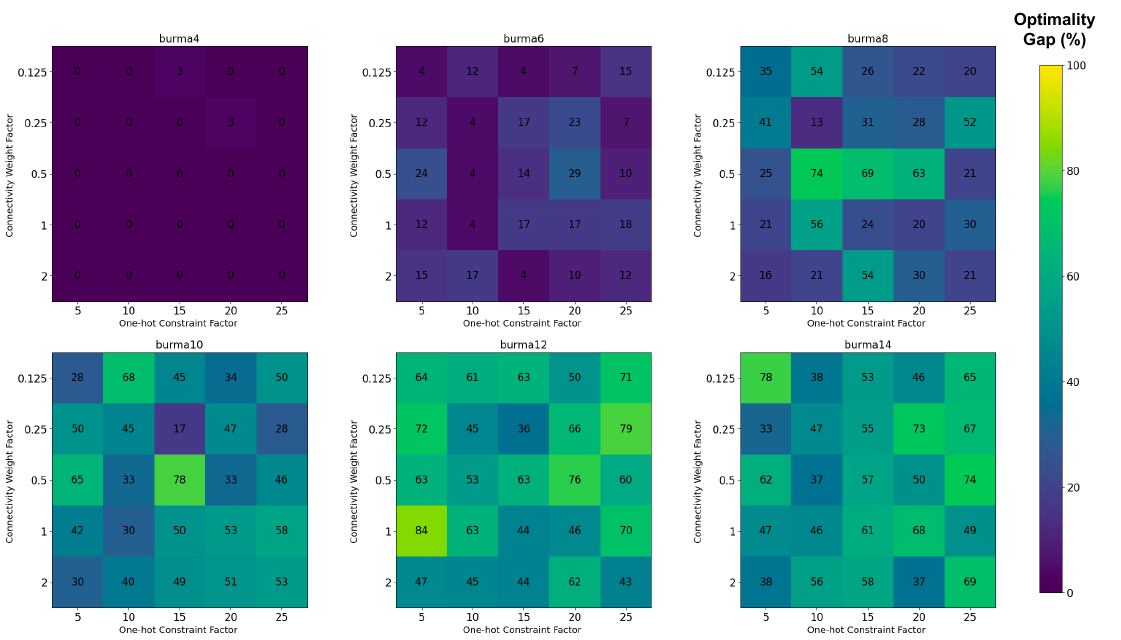}
\par\end{centering}
\caption{\label{fig:hyperparam_tsp} Parameter Search across tsp problem instances for Ising framework. It searches for the optimal value of the connectivity weight factor and one-hot constraint factor to minimize the optimality gap. The optimality gap quantifies the relative difference in tour cost obtained by the Ising machines compared with the optimal tour achieved using the Lin-Kernighan Heuristic} \cite{lin1973heuristic}.

\end{figure*}
\clearpage

\begin{figure*}[h!]
\begin{centering}
\includegraphics[width=\linewidth]{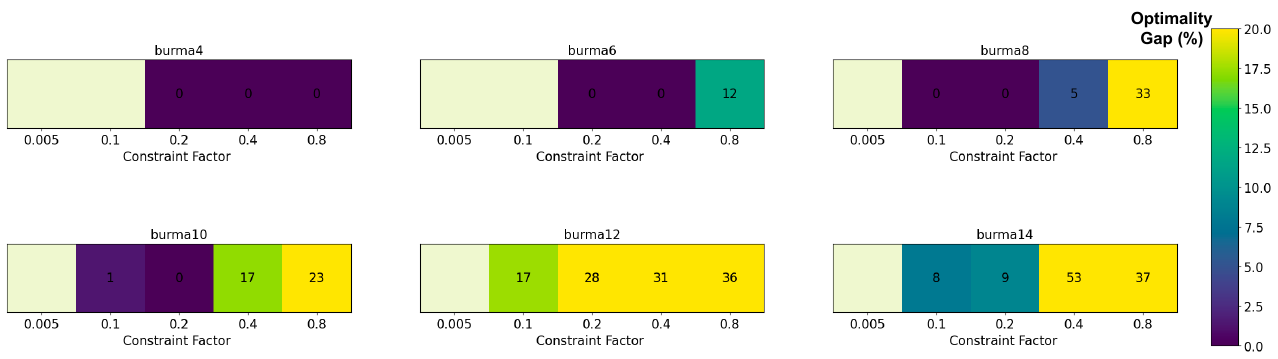}
\par\end{centering}
\caption{\label{fig:hyperparam_tsp_vec} Parameter ($wt$) Search across tsp problem instances for vectorized framework.}
\end{figure*}
\clearpage

\begin{figure*}[h!]
\begin{centering}
\includegraphics[width=\linewidth]{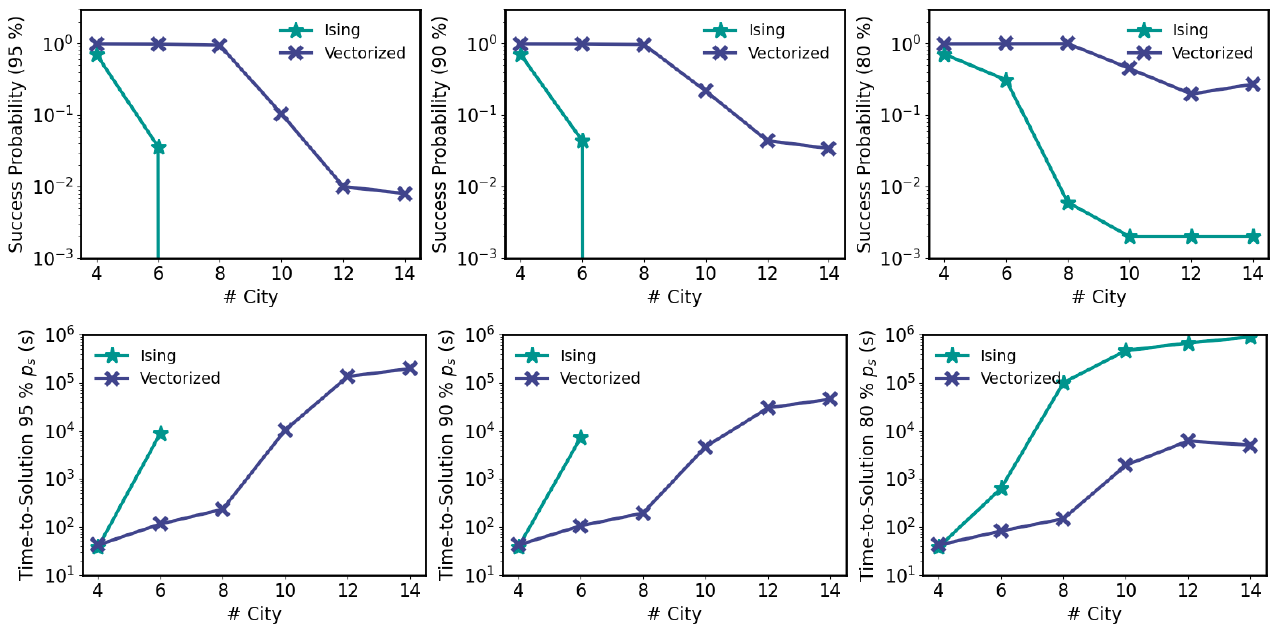}
\par\end{centering}
\caption{\label{fig:tsp_ising_vec} Success probability and TTS metric for 99 $\%$ success comparing the solution quality and efficiency of Ising and vectorized mapping framework for TSP problem instances upto 14 cities.}
\end{figure*}

\clearpage

\begin{figure*}[h!]
\begin{centering}
\includegraphics[width=0.4\linewidth]{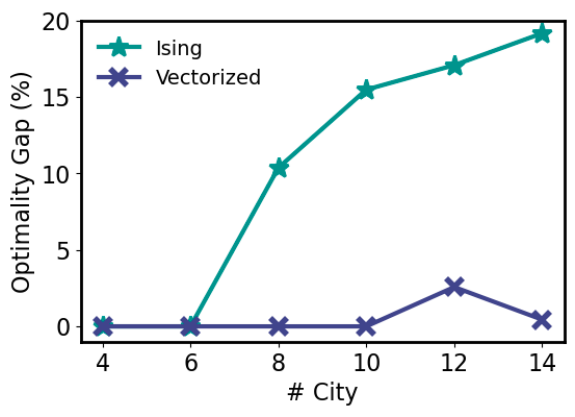}
\par\end{centering}
\caption{\label{fig:optimality_gap} Best Optimality gap achieved by Ising and vectorized mapping framework for TSP problem instances upto 14 cities.}
\end{figure*}

\clearpage

\begin{figure*}[h!]
\begin{centering}
\includegraphics[width=1\linewidth]{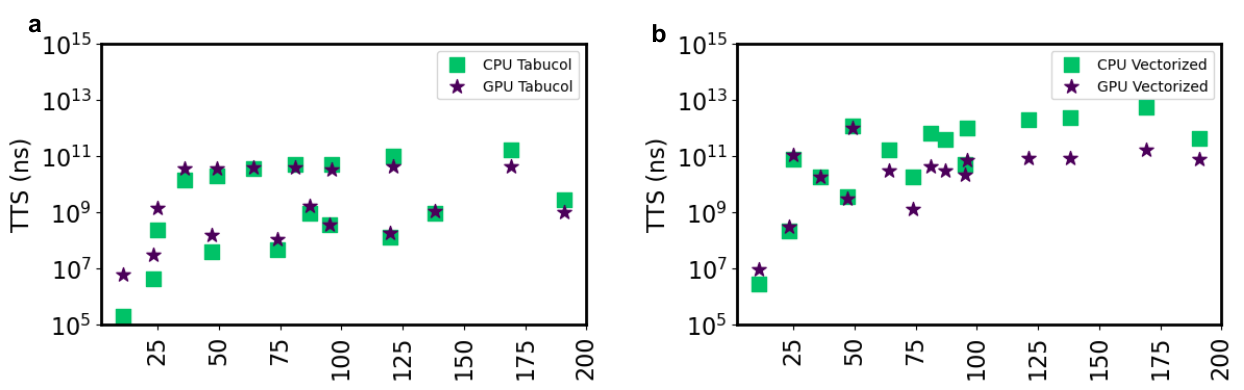}
\par\end{centering}
\caption{\label{fig:cpu_gpu_tabu_vec} Performance comparison between CPU and GPU implementation of Tabucol heuristics and Vectorized framework.}
\end{figure*}



\end{document}